\documentclass[printer]{aa}  

\usepackage{graphicx}
\usepackage{txfonts}
\begin{document}

   \title{Kinetic Electrostatic Waves and their Association with Current Structures in the Solar Wind}
   \titlerunning{Kinetic Electrostatic Waves and Current Structures}

\author{D. B. Graham\inst{1}
\and Yu. V. Khotyaintsev\inst{1}
\and A. Vaivads\inst{2}
\and N. J. T. Edberg\inst{1}
\and A. I. Eriksson\inst{1}
\and E. Johansson\inst{1}
\and L. Sorriso-Valvo\inst{1}
\and M. Maksimovic\inst{3}
\and J. Sou\v{c}ek\inst{4}
\and D. P\'{i}\v{s}a\inst{4}
\and S. D. Bale\inst{5,6}
\and T. Chust\inst{7}
\and M. Kretzschmar\inst{8,9}
\and V. Krasnoselskikh\inst{8,5}
\and E. Lorf\`evre\inst{10}
\and D. Plettemeier\inst{11}
\and M. Steller\inst{12}
\and \v{S}. \v{S}tver\'ak\inst{13}
\and P. Tr\'avn\'i\v{c}ek\inst{5,4}
\and A. Vecchio\inst{3,14}
\and T. S. Horbury\inst{15}
\and H. O'Brien\inst{15}
\and V. Evans\inst{15}
\and V. Angelini\inst{15}
          }
   \institute{Swedish Institute of Space Physics (IRF), Uppsala 75121, Sweden\\
              \email{dgraham@irfu.se}
\and Division of Space and Plasma Physics, School of Electrical Engineering and Computer Science,
KTH Royal Institute of Technology, Stockholm 11428, Sweden
\and LESIA, Observatoire de Paris, Universit\'e PSL, CNRS, Sorbonne Universit\'e, Univ. Paris Diderot, Sorbonne Paris Cit\'e, 5 place Jules Janssen, 92195 Meudon, France
\and Institute of Atmospheric Physics of the Czech Academy of Sciences, Prague, Czechia
\and
Space Sciences Laboratory, University of California, Berkeley, CA, USA
\and
Physics Department, University of California, Berkeley, CA, USA
\and LPP, CNRS, Ecole Polytechnique, Sorbonne Universit\'{e}, Observatoire de Paris, Universit\'{e} Paris-Saclay, Palaiseau, Paris, France
\and LPC2E, CNRS, 3A avenue de la Recherche Scientifique, Orl\'eans, France
\and Universit\'e d'Orl\'eans, Orl\'eans, France
\and CNES, 18 Avenue Edouard Belin, 31400 Toulouse, France
\and Technische Universität Dresden, Helmholtz Str. 10, D-01187 Dresden, Germany
\and Space Research Institute, Austrian Academy of Sciences, Graz, Austria
\and Astronomical Institute of the Czech Academy of Sciences, Prague, Czechia
\and Radboud Radio Lab, Department of Astrophysics, Radboud University, Nijmegen, The Netherlands
\and Imperial College London, South Kensington Campus, London SW7 2AZ, UK
             }
   
   \date{Received September 15, 1996; accepted March 16, 1997}

 
  \abstract
   {A variety of kinetic electrostatic and electromagnetic waves develop in the solar wind. The relationship between these waves and larger-scale structures, such as current sheets and ongoing turbulence remain a topic of investigation. Similarly, the instabilities producing ion-acoustic waves in the solar wind remains an open question.}
   {The goals of this paper are to investigate kinetic electrostatic Langmuir and ion-acoustic waves in the solar wind at 0.5 AU and determine whether current sheets and associated streaming instabilities can produce the observed waves. The relationship between these waves and currents observed in the solar wind is investigated statistically.}
   {Solar Orbiter's  Radio and Plasma Waves (RPW) instrument suite provides high-resolution snapshots of the fluctuating electric field. The Low Frequency Receiver (LFR) resolves the waveforms of ion-acoustic waves and the
   Time Domain Sampler (TDS) resolves the waveforms of both ion-acoustic and Langmuir waves. Using these waveform data we determine when these waves are observed in relation to current structures in the solar wind, estimated from the background magnetic field.}
   {Langmuir and ion-acoustic waves are frequently observed in the solar wind. Ion-acoustic waves are observed about $1 \%$ of the time at 0.5 AU. The waves are more likely to be observed in regions of enhanced currents. However, the waves typically do not occur at current structures themselves. The observed currents in the solar wind are too small to drive instability by the relative drift between single ion and electron populations. When multi-component ion and/or electron distributions are present the observed currents may be sufficient for instability. Ion beams are the most plausible source of ion-acoustic waves in the solar wind. The spacecraft potential is confirmed to be a reliable probe of the background electron density by comparing the peak frequencies of Langmuir waves with the plasma frequency calculated from the spacecraft potential. }
   {}


   \maketitle
%

\section{Introduction}
The solar wind is weakly collisional, and as a result non-Maxwellian and complex electron and ion distributions can develop \cite[e.g.,][]{marsch2006}. Such distributions may generate kinetic plasma waves in the solar wind. In the solar wind many different types of electrostatic and electromagnetic plasma waves have been observed. These include Langmuir waves, ion-acoustic waves, electrostatic solitary waves (ESWs), whistler waves, lower hybrid waves, and Alfv\'en waves \cite[e.g.,][]{schwartz1980}. 

Early wave observations in the solar wind found electrostatic oscillations at a few kHz \cite[]{gurnett1977}. These oscillations were interpreted as ion-acoustic waves Doppler shifted to higher frequencies by the solar wind flow \cite[]{gurnett1977,gurnett1978,gurnett1979}. 
Additionally, nonlinear structures such as ESWs have been observed in the solar wind \cite[]{mangeney1999,malaspina2013}. Various streaming instabilities have been proposed to generate ion-acoustic waves. These include the electron-ion streaming instability \cite[]{bernstein1960}, the ion-ion-acoustic instability \cite[]{lemons1979,gary1987}, and electron-electron-ion streaming instability \cite[]{lapuerta2002,norgren2015}. Additionally, temperature gradients are also a possible source of ion-acoustic waves \cite[]{allan1974}. 

In the solar wind, Langmuir waves are typically observed in the source regions of type II and type III solar radio bursts \cite[]{pulupa2008,graham2013,graham2015}, as well as in planetary electron foreshock regions \cite[e.g.,][]{filbert1979}. In these regions Langmuir waves can readily reach large amplitudes ($\gtrsim 10$~mV~m$^{-1}$ in the solar wind at 1 AU) \cite[]{graham2014,graham2015}. These waves are generated by the bump-on-tail instability, and undergo linear or nonlinear processes to generate radio waves at the local electron plasma frequency and the second harmonic \cite[]{ginzburg1958,melrose1980,kellogg1980}. 
However, Langmuir waves have also been observed in the solar wind, which is not associated with type II and III source regions. For example, they have been observed in relation to magnetic holes \cite[]{lin1996,braind2010}. 
Similarly, Langmuir waves have been reported in association with solar wind current sheets \cite[]{lin1996} and magnetic reconnection \cite[]{huttunen2007}. 

The goal of this paper is to investigate the relation between kinetic ion-acoustic and Langmuir waves and currents observed in the solar wind. Our results show that the current densities estimated in the solar wind are too small to generate ion-acoustic waves by an electron-ion streaming instability, consisting of single electron and ion distributions. Instead multi-component electron or ion distributions are required for instability. Statistically we find that the ion-acoustic waves are weakly correlated with current structures in the solar wind. Additionally, we compare the observed frequency of Langmuir waves to the electron plasma frequency estimated from the spacecraft potential, to confirm the reliability of the spacecraft potential as a probe of the background electron number density \cite[]{khotyaintsev2021}.

The outline of this paper is as follows: In section \ref{instruments} we introduce the instruments and data used in this study. Section \ref{theory} provides a brief overview of the streaming instabilities than can generate ion-acoustic-like waves. In Section \ref{kineticwaves} we present observations of ion-acoustic and Langmuir waves. In Section \ref{Lwaves} we also compare the Langmuir wave frequencies with the electron plasma frequency calculated from the spacecraft potential to confirm the reliability of the spacecraft potential as a density probe. In section \ref{swcurrents} we investigate the relationship between electrostatic waves and currents observed in the solar. In section \ref{conclusions} we state the conclusions. 

\section{Instruments and Data} \label{instruments}
In this paper we use fields data from the Solar Orbiter spacecraft. We use electric field and potential data from the Low Frequency Receiver (LFR) and electric field data from the Time Domain Sampler (TDS) receiver of the Radio and Plasma Waves (RPW) instrument suite \cite[]{maksimovic2020} and magnetic field data from the Solar Orbiter Magnetometer \cite[]{horbury2020}. The Magnetometer data are nominally sampled at $8$~Hz. The LFR receiver provides continuous data, as well as high-frequency waveform snapshots captured at regular intervals. Snapshots are recorded at three different sampling rates: $256$~Hz, $4$~kHz, and $25$~kHz. Only the $25$~kHz can reliably 
resolve ion-acoustic waves, which typically have frequencies ranging from a few hundred Hz to several kHz, so this is the only sampling rate used in this paper. With these snapshots the maximum resolvable frequency is $\approx 10$~kHz. We note that the frequency of ion-acoustic waves can exceed $10$~kHz in some rare cases. 
The TDS receiver provides very high-resolution electric field snapshots, which are triggered by wave activity in the solar wind. These snapshots resolve ion-acoustic waves and Langmuir waves at the local electron plasma frequency $f_{pe}$.
All vector data are presented in the Spacecraft Reference Frame (SRF) coordinates, where x is Sunward, y is close to anti-aligned with the tangential direction, and z is close to the normal direction, unless otherwise stated. 
The probe-to-spacecraft potential $V_{psp}$ and potential differences between the probes are calculated using the three $6.5$~m monopole electric antennas. By combining the measurements from the three antennas we calculate the two-dimensional electric field in the SRF y- and z-directions. The electric field ${\bf E}$ is calculated using 
\begin{equation}
E_y = -\frac{V_{23}}{L_y}, E_z = -\frac{(V_{12} +0.5  V_{23})}{L_z},
\label{Eyzcal}
\end{equation}
where $V_{23}$ and $V_{12}$ are the potential differences between probes 2 and 3 and 1 and 2, respectively, and $L_y = 6.99$~m and $L_z = 6.97$~m are the effective lengths in the y- and z-directions used in this paper. 
We apply this same calibration to both LFR and TDS potential data. We note that the TDS receiver channels were configured in three different ways to telemeter different combinations of probe potentials or potential differences between probe pairs over June 2020. For each case we calculate $V_{12}$ and $V_{23}$ from the 
TDS receiver channels to compute ${\bf E}$. 

\begin{figure}[htbp!]
\begin{center}
\includegraphics[width=80mm, height=90mm]{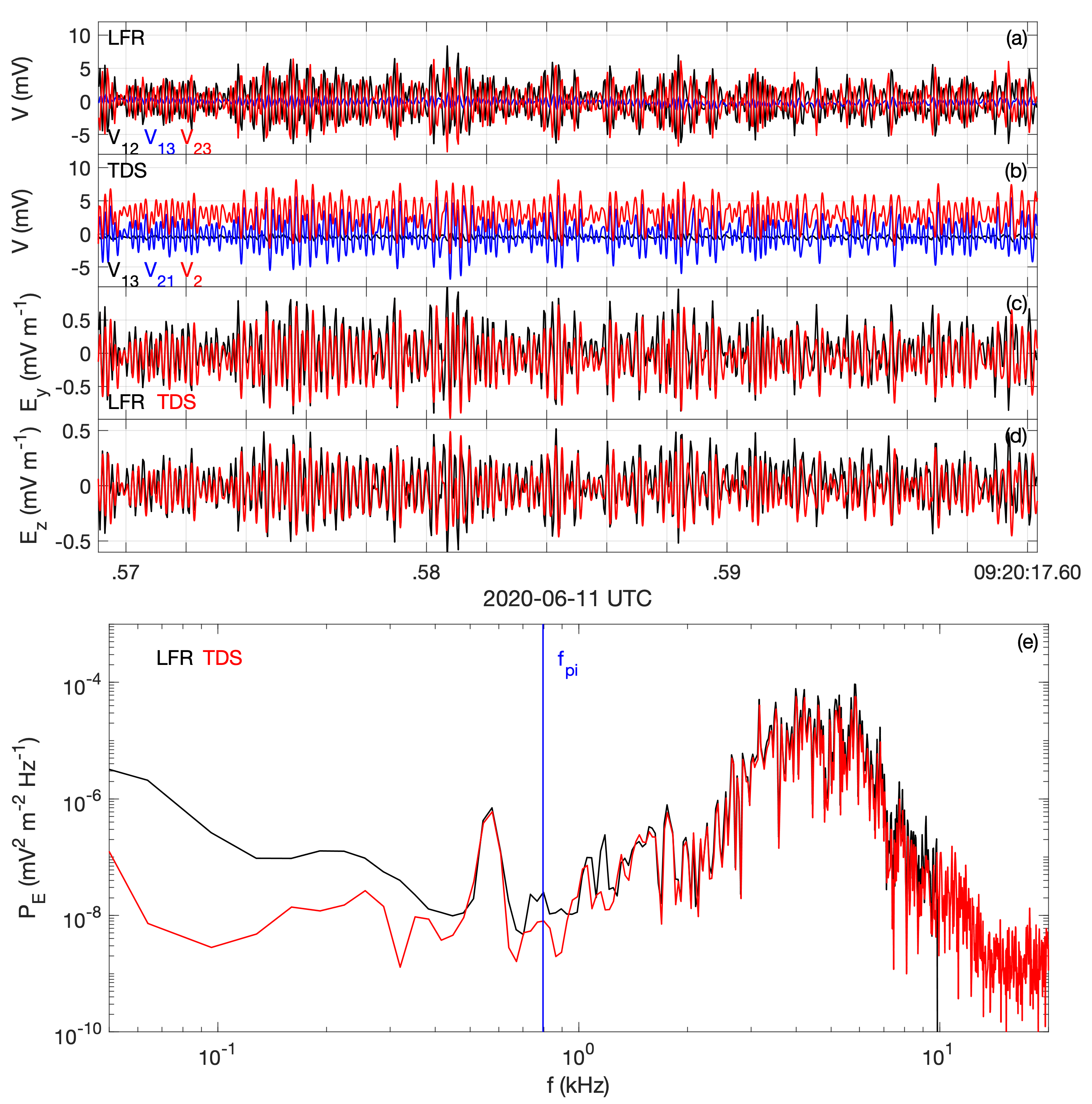}
\caption{An ion-acoustic wave simultaneously observed by the LFR and TDS receivers. (a) Potentials $V_{12}$ (black), $V_{13}$ (red), and $V_{23}$ (blue) from LFR. (b) Potentials $V_{13}$ (black), $V_{21}$ (red), and $V_{2}$ (blue) from TDS. (c) $E_y$ computed from the LFR (black) and TDS receivers. (d) $E_z$ computed from the LFR (black) and TDS receivers. (e) Power spectra of ${\bf E}$ from LFR (black) and TDS (red). }
\label{LFRTDScomp}
\end{center}
\end{figure}

As an example of the calculation of ${\bf E}$, Figure \ref{LFRTDScomp} shows an ion-acoustic wave observed when LFR and TDS snapshots were recorded simultaneously. In Figure \ref{LFRTDScomp} we plot the original potential data and the calculated ${\bf E}$. Figure \ref{LFRTDScomp}a shows the potentials $V_{12}$, $V_{13}$, and $V_{23}$ from LFR and Figure \ref{LFRTDScomp}b shows the potentials $V_{13}$, $V_{21}$, and $V_{2}$ from TDS. Note that the TDS potentials differ from LFR and consist of dipole and monopole components. 
The y- and z-components of ${\bf E}$ from LFR and TDS are shown in Figures \ref{LFRTDScomp}c and \ref{LFRTDScomp}d, respectively. The electric field from LFR and TDS are almost identical, as expected. Similarly, 
the power spectra of ${\bf E}$,  shown in Figure \ref{LFRTDScomp}e, from LFR and TDS are almost identical. 
Similar results are found for other events where LFR and TDS capture snapshots simultaneously (not shown).

We also use $V_{psp}$ from LFR to calculate the local solar wind electron density $n_{e,SC}$, which is calibrated by comparing $V_{psp}$ with the plasma line identified from quasi-thermal noise. The details of the calibration can be found in \cite[]{khotyaintsev2021}.
In section \ref{Lwaves} we compare $n_e$ calculated with the frequencies of Langmuir waves $n_{e,pk}$ observed by TDS to assess the reliability of $n_{e,SC}$. 

In this paper, we investigate ion-acoustic and Langmuir waves observed over June 2020, when Solar Orbiter was close to its first perihelion. At this time 
Solar Orbiter was located at $\sim$0.5~AU from the Sun. 

\section{Theory} \label{theory}
In this section we consider the theory of electrostatic kinetic instabilities, which can generate ion-acoustic waves in the solar wind. To model the instability we use the kinetic unmagnetized electrostatic dispersion equation
\begin{equation}
0 = 1 - \sum_{j} \frac{\omega_{pj}^2}{k^2 v_j^2} Z' \left( \zeta_j \right),
\label{eseq}
\end{equation}
where $\omega_p$ is the angular plasma frequency, $k$ is the wave number, $v = \sqrt{2 k_B T/m}$, $\zeta = (\omega - k V_j)/k v_j$, $V$ is the bulk speed aligned with $k$, $k_B$ is Boltzmann's constant, $T$ is the temperature, $m$ is the particle mass, $Z' = -2 [1 + \zeta Z(\zeta)]$ is the derivative of the plasma dispersion function $Z$ \cite[]{fried1961}, and the subscripts $j$ refer to the different particle species. Equation (\ref{eseq}) assumes the distributions are Maxwellian. We consider the following three cases: 
(1) One electron and one ion component, (2) two electron components and one ion component, and (3) 
one electron component and two ion components. 
In each case the relative drift between difference components is the cause of instability. 

\begin{figure}[htbp!]
\begin{center}
\includegraphics[width=90mm, height=120mm]{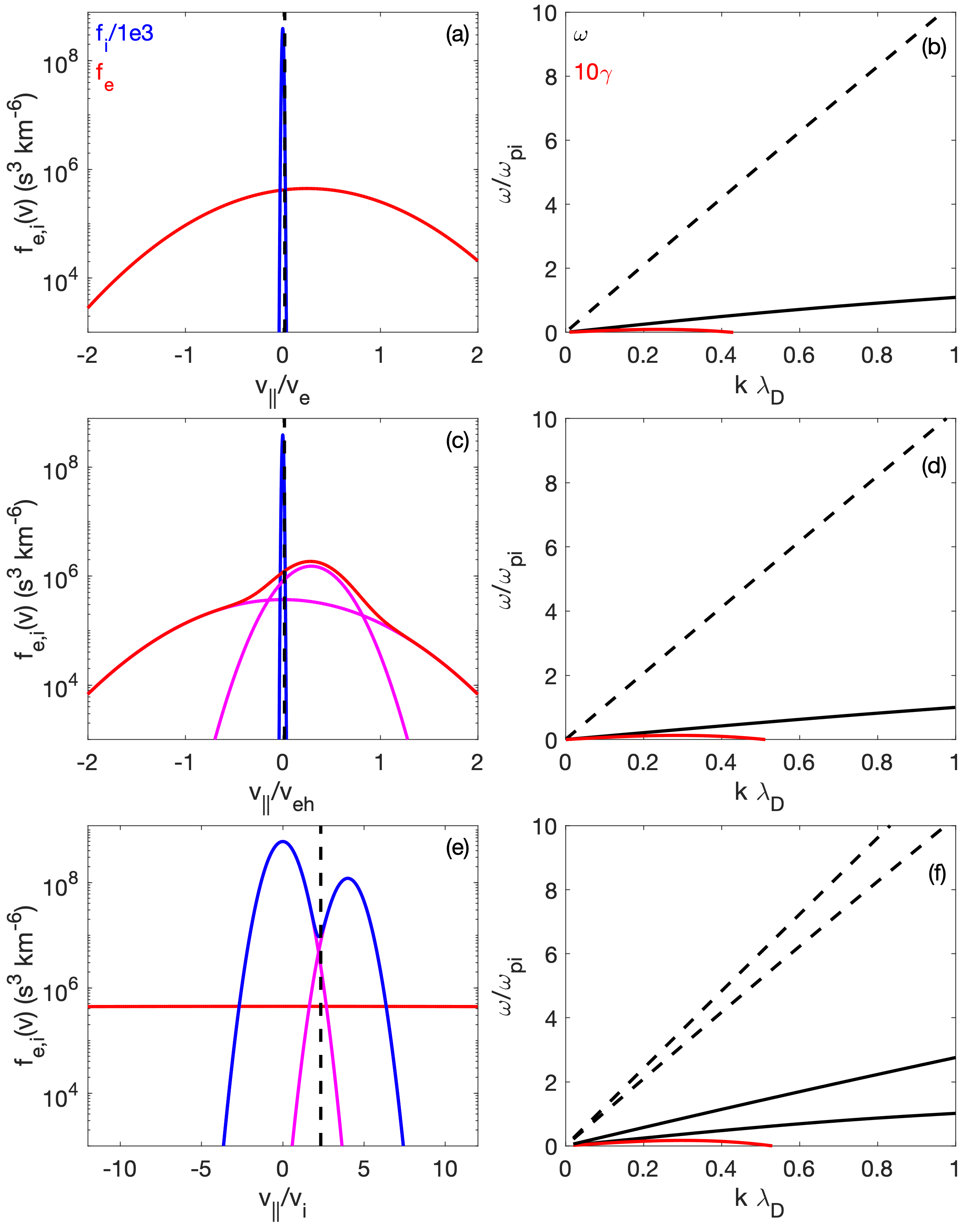}
\caption{Parallel streaming instabilities and the dispersion relations of ion-acoustic waves. (a) The electron and ion distribution functions of the electron-ion streaming instability. For the electron distribution we use $n_e = 30$~cm$^{-3}$, $T_e = 15$~eV and $V_e = 0.25 v_e$ and for the ion distribution we use $n_i = 30$~cm$^{-3}$, $T_i = 3$~eV. (c) The distribution for the electron-electron-ion streaming instability, consisting of core electrons and ions, and an electron beam. For core electrons we use $n_e = 25$~cm$^{-3}$ and $T_{e} = 15$~eV, for core ions $n_i = 30$~cm$^{-3}$ and $T_i = 3$~eV, and beam electrons $n_b = 5$~cm$^{-3}$, $T_b = 2$~eV, and $V_b = 0.8 v_b$. (e) The distribution for the ion beam-driven instability, consisting of core electrons and an ion beam. For core electrons we use $n_e = 30$~cm$^{-3}$, $T_e = 15$~eV, for core ions $n_i = 25$~cm$^{-3}$ and $T_i = 2$~eV, and beam ions $n_b = 5$~cm$^{-3}$, $T_b = 2$~eV, and $V_b = 4 v_b$. Panels (b), (d), and (f) show the dispersion relations calculated from the distributions in panels (a), 
(c), and (e), respectively. The solid black lines show the dispersion relations in the plasma rest frame or $\theta_{Bx} = 90^{\circ}$, while the dashed black lines show the dispersion relations Doppler shifted due to the solar wind flow past the spacecraft for $\theta_{Bx} = 0^{\circ}$ and $V_{sw} = 350$~km~s$^{-1}$. The red curves show the growth rate $\gamma$. In panel (f) the higher-frequency dispersion relation is stable $\gamma < 0$ for all $k$, while the lower-frequency mode is unstable. }
\label{minst} 
\end{center}
\end{figure}

Figure \ref{minst} shows an example of each of these instabilities. We have chosen drift parameters marginally larger than the threshold required for growth to assess the current densities associated with each instability. 
In each case the total number density is $30$~cm$^{-3}$, which is the median density measured over June 2020. Figure \ref{minst}a shows the electron and ion distributions with a relative drift (case 1) and the associated unstable mode (Figure \ref{minst}b). The unstable mode, real frequency $\omega$ and growth rate $\gamma$, are shown, as well as the Doppler shifted dispersion relation (black dashed line) when ${\bf k}$ and the slow wind velocity ${\bf V}_{sw}$ are aligned for $V_{sw} = 350$~km~s$^{-1}$. We have used $T_e = 15$~eV, $T_i = 3$~eV, and use an electron drift speed of $0.25 v_e$. Without Doppler shift the dispersion relation has frequencies ranging from $0$ at $k \lambda_D = 0$ 
to $\sim \omega_{pi}$ at $k \lambda_D = 1$, where $\omega_{pi}$ is the ion plasma frequency and $\lambda_D$ is the Debye length. For the Doppler shifted dispersion relation we find that the frequency is substantially larger than $\omega$ in the rest frame, thus the linear dispersion relation becomes $\omega \approx k V_{sw}$, i.e., the frequency increases approximately linearly with $k$. This is perhaps unsurprising, since the ion-acoustic speed is predicted to be $C_S \approx 50$~km~s$^{-1}$ for the conditions used in the model, and thus $C_S \ll V_{sw}$. Therefore, the observed wave frequency will typically be substantially larger than the wave frequency in the plasma rest frame, except when ${\bf B}$ and 
${\bf V}_{sw}$ is close to perpendicular. 

In Figure \ref{minst}c we show a distribution consisting of a single ion population, a core electron population, and a dense electron beam that is $17 \%$ of the total electron density (case 2). This distribution is unstable to the electron-electron-ion instability. For these conditions the source of instability is the interaction between the electron beam and the stationary ion population. The resulting dispersion relation is almost the same as in Figure \ref{minst}b. As the beam speed is increased the beam electrons will interact with the core electrons instead of the ions, generating beam-mode waves or electron-acoustic waves, with significantly higher phase speeds and frequencies. Thus, distributions of this form can be unstable to ion-acoustic waves or higher-frequency electrostatic waves. 

In Figure \ref{minst}e we consider the case of two ion populations (case 3), a core ion population and an ion beam, and a single electron population. For this distributions two modes in the ion plasma frequency range are found (Figure \ref{minst}f). The higher frequency mode is the ion-acoustic wave.
This mode is stable for the conditions used in Figure \ref{minst}e (see caption). The lower-frequency mode is the ion-ion-acoustic mode (also called the ion beam-driven mode), which is unstable due to the interaction between the ion beam and core ion population. The dispersion relation is very similar to previous examples.

To provide an indication of the current density $J$ required for instability, we compute $J$ associated with the distributions in Figure \ref{minst}, using $J = \sum_j e_j n_j V_j$, where $e_j$ is the charge of each particle species. The current densities are $J = 2.8 \times 10^{3}$~nA~m$^{-2}$, $5.4 \times 10^{2}$~nA~m$^{-2}$, and $60$~nA~m$^{-2}$ for the distributions in panels (a), (c), and (e), respectively. The value of $J$ required for instability due to the relative drift between ions and electrons in Figure \ref{minst}a is extremely large, and unlikely to occur at solar wind current sheets. When multiple electron and ion distributions are present, $J$ required for instability can be substantially reduced. We also note that for either multi-component electron and ion distributions it is possible for instability to occur for $J = 0$, for example when the core and beam electrons have opposite bulk speeds relative to the ions for the electron-electron-ion instability, and \textit{vice versa} for the ion-beam driven case. 

In summary, we propose three streaming instabilities as possible sources of ion-acoustic waves. These three instabilities produce nearly identical dispersion relations, so they cannot be distinguished from each other based on wave observations alone. Of these, the relative drift between a single ion and electron distribution is the most unlikely as this requires extremely large currents, which are unlikely to be observed in the solar wind. By allowing more complex distributions consisting of either multiple electron or ion components, the currents required for instability can be substantially reduced, and can theoretically occur for $J = 0$. 
This makes these instabilities more plausible as sources of ion-acoustic waves in the solar wind. 

\section{Kinetic waves} \label{kineticwaves}
\subsection{Ion-acoustic waves}
\begin{figure*}[htbp!]
\begin{center}
\includegraphics[width=140mm, height=120mm]{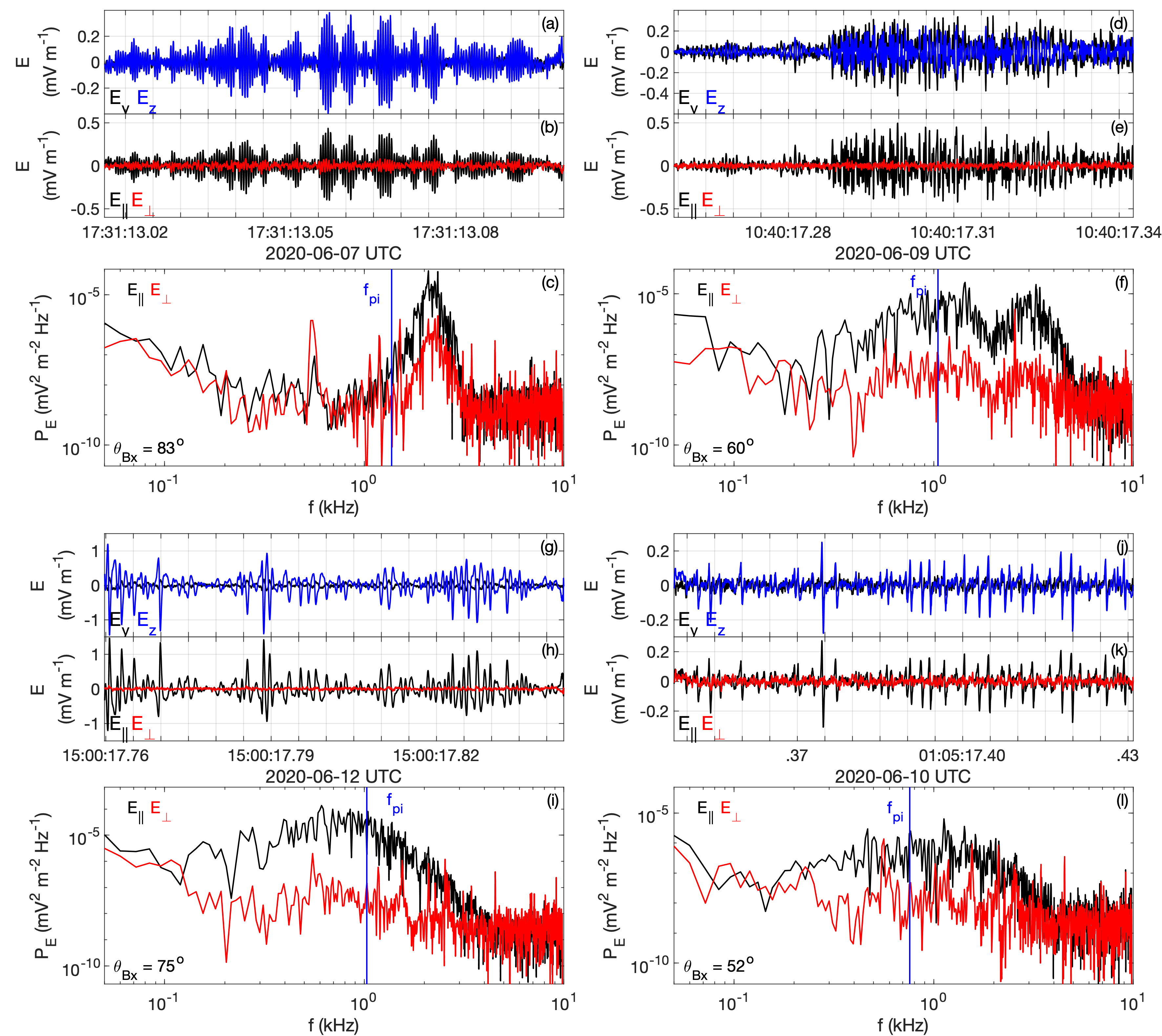}
\caption{Four examples of ion-acoustic-like waves seen by LFR, shown in panels (a)--(c), (d)--(f), (g)--(i), and 
(j)--(l), respectively. For each event we plot $E_y$ and $E_z$ in panels (a), (d), (g), and (j), 
$E_{\parallel}$ and $E_{\perp}$ in (b), (e), (h), and (k), and the power spectra of $E_{\parallel}$ and $E_{\perp}$ 
in panels (c), (f), (i), and (l). In panels (c), (f), (i), and (l) the vertical blue line indicates the local ion plasma frequency $f_{pi}$ calculated from $n_{e,SC}$. }
\label{LFRexamples}
\end{center}
\end{figure*}

In this section we show the observation of ion-acoustic waves and show some example waveforms seen by LFR. 
For the purposes of this paper we include all electrostatic waves at $\sim f_{pi}$ and above. This includes nonlinear structures, such as electrostatic solitary waves. Figure \ref{LFRexamples} shows four examples of the types of waves observed by LFR. For each case we plot the time series of $E_y$ and $E_z$, the time series of $E_{\parallel}$ and $E_{\perp}$, and the power spectra of $E_{\parallel}$ and $E_{\perp}$. 
Since we only have two components of ${\bf E}$, we define $E_{\parallel}$ as the component aligned with ${\bf B}$ in the y-z plane, while $E_{\perp}$ is perpendicular to ${\bf B}$. Note that $E_{\perp}$ is always perpendicular to ${\bf B}$, while $E_{\parallel}$ contains both parallel and perpendicular components of ${\bf E}$, depending on the angle $\theta_{Bx}$ between ${\bf B}$ and the SRF x-direction. In the case that the wave electric field is aligned with ${\bf B}$, we expect $E_{\parallel} \gg E_{\perp}$ with $E_{\parallel}$ underestimating the true value by $\sin^{-1}\theta_{Bx}$. In all four cases we find that $E_{\parallel} \gg E_{\perp}$, consistent with ${\bf E}$ being aligned with ${\bf B}$. This is generally found for ion-acoustic waves seen in the LFR and TDS data. The statistical properties of ion-acoustic waves are investigated by \cite[]{pisa2021}.

The first example, Figures \ref{LFRexamples}a--\ref{LFRexamples}c, shows a waveform with relatively sinusoidal fluctuations in $E_{\parallel}$ and has frequencies just above the local ion plasma frequency $f_{pi}$. Compared 
with the other examples, the power occurs over a relatively small range of frequencies, which results in periodic 
fluctuations. The second example, Figures \ref{LFRexamples}d--\ref{LFRexamples}f, shows a more complex waveform. The waveform is no longer clearly periodic, but rather the fluctuations are more complex. 
The power spectrum (Figures \ref{LFRexamples}f) is much broader than in the first example, and has two spectral peaks near $f_{pi}\sim 1$~kHz and $\sim 3$~kHz. The third example, Figures \ref{LFRexamples}g--\ref{LFRexamples}i, shows a non-sinusoidal waveform, with a broad spectral peak centered around $f_{pi}$. 
The non-sinusoidal nature of the waves might suggest that nonlinear processes are occurring, such as the formation of solitary structures. 
The final example, Figures \ref{LFRexamples}j--\ref{LFRexamples}l, shows a waveform corresponding to a series of electrostatic solitary waves (ESWs). 
The ESWs are characterized by bipolar fluctuations in $E_{\parallel}$, and typically correspond to electron phase-space holes. The ESWs are regularly spaced. 
For all ESWs there is a positive $E_{\parallel}$ followed by a negative $E_{\parallel}$, which suggests that the ESWs all propagate in the same direction. 
The power spectrum (Figure \ref{LFRexamples}l) exhibits a broad range of frequencies near $f_{pi}$ and has a similar shape to the spectrum of the third example. 
In each case we find that wave power extends well above $f_{pi}$ and the spectral peaks typically occur above $f_{pi}$. This is due to Doppler shift in the solar wind, as illustrated in Figure \ref{minst}.

We search for ion-acoustic waves in LFR and TDS data. 
For the TDS data we use the following criteria: (1) The maximum electric field satisfies $E_{max} > 0.05$~mV~m$^{-1}$, and (2) the waves have frequencies $f$ greater than $200$~Hz. 
For the purposes of this search we considered all waveforms exemplified in Figure \ref{LFRexamples}, including sinusoidal waves, non-sinusoidal fluctuations, and ESWs. 
From TDS we identify 2553 waveform snapshots of ion-acoustic waves, 
corresponding to $60~\%$ of the TDS snapshots captured in June 2020. Since the TDS snapshots are triggered by wave activity they are far more likely to be observed by TDS than LFR.
For LFR snapshot data we use the same search criteria, except $E_{max} > 0.08$~mV~m$^{-1}$, due to the higher background noise level on LFR. From the data in June 2020 we identify 423 waveform snapshots of ion-acoustic waves, which corresponds to $1.3~\%$ of all recorded snapshots. Since the LFR snapshots are taken regularly, rather than triggered by wave activity, this percentage provides an indication of how common ion-acoustic waves are in the solar wind at $0.5$~AU. 

\subsection{Langmuir waves} \label{Lwaves}
Langmuir waves are electrostatic waves observed near the electron plasma frequency $f_{pe}$. The waves are typically generated by electron beams. In this section we look at the properties of Langmuir waves in the solar wind and compare the observed Langmuir wave frequencies to $f_{pe,SC}$, where $f_{pe,SC}$ is the electron plasma calculated from $n_{e,SC}$. 

\begin{figure*}[htbp!]
\begin{center}
\includegraphics[width=150mm, height=75mm]{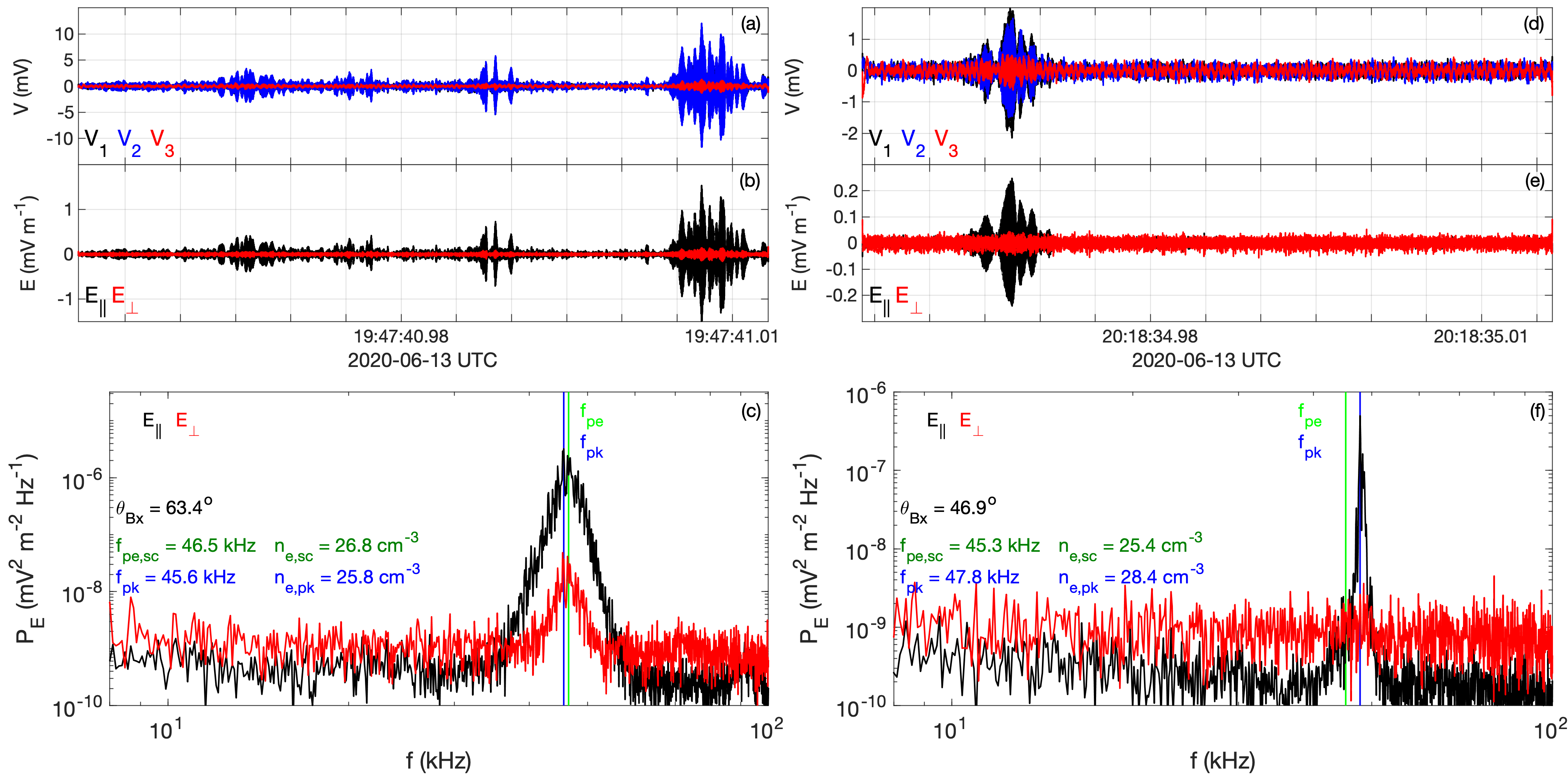}
\caption{Two examples of Langmuir waves observed by TDS on 13 June 2020. Panels (a)--(c) show a Langmuir or beam-mode wave with a relatively broad spectral peak, while panels (d)--(f) show a typical example of Langmuir waves with a narrow spectral peak. Panels (a) and (d) show the waveforms of the three potentials from TDS. For these cases TDS provide three monopole potentials $V_{1-3}$ from each antenna. Panels (b) and (e) show the waveforms of $E_{\parallel}$ and $E_{\perp}$. Panels (c) and (f) show the power spectra of $E_{\parallel}$ and $E_{\perp}$. The blue and green vertical lines indicate $f_{pe}$ calculated from $n_{e,SC}$ and the frequency at which the power peaks $f_{pk}$. }
\label{Leg}
\end{center}
\end{figure*}

Figure \ref{Leg} shows two examples of Langmuir waves seen by TDS. In Figure \ref{Leg} we plot the potentials measured by TDS, $V_1$, $V_2$, and $V_3$, the electric field in field-aligned coordinates, and power spectra of $E_{\parallel}$ and $E_{\perp}$. For both examples we find that $E_{\parallel} \gg E_{\perp}$. This is typically found for Langmuir waves seen in the solar wind by Solar Orbiter (not associated with type II or III source regions). The first example, Figures \ref{Leg}a-\ref{Leg}c, shows a relatively broad spectral peak centered around $f_{pe}$. This broad spectral peak may correspond to beam-mode waves rather than Langmuir waves. We note that beam-mode waves can have frequencies both below and above $f_{pe}$ \cite[]{fuselier1985,soucek2019}. The waveform shows that the wave is bursty, with a series of localized wave packets. 
The second example, Figures \ref{Leg}d--\ref{Leg}f, shows a Langmuir wave with a very narrow spectral peak, just above $f_{pe,SC}$. In this case, the waveform is quite localized and is characterized by $E_{\parallel} \gg E_{\perp}$. These characteristic are typical of Langmuir waves observed in the solar wind. From the peak in power we can estimate the local $f_{pe}$, and hence $n_e$, by assuming that the frequency of peak Langmuir power corresponds to $f_{pe}$. The frequencies $f_{pe,SC}$ and $f_{pk}$ are shown in Figures \ref{Leg}c and \ref{Leg}f. For the first example we find that $f_{pe,SC} \approx f_{pk}$, while for the second example $f_{pk}$ exceeds $f_{pe,SC}$ by $2.5$~kHz, which could suggest that 
$f_{pe,SC}$ is underestimated. 

\begin{figure*}[htbp!]
\begin{center}
\includegraphics[width=140mm, height=120mm]{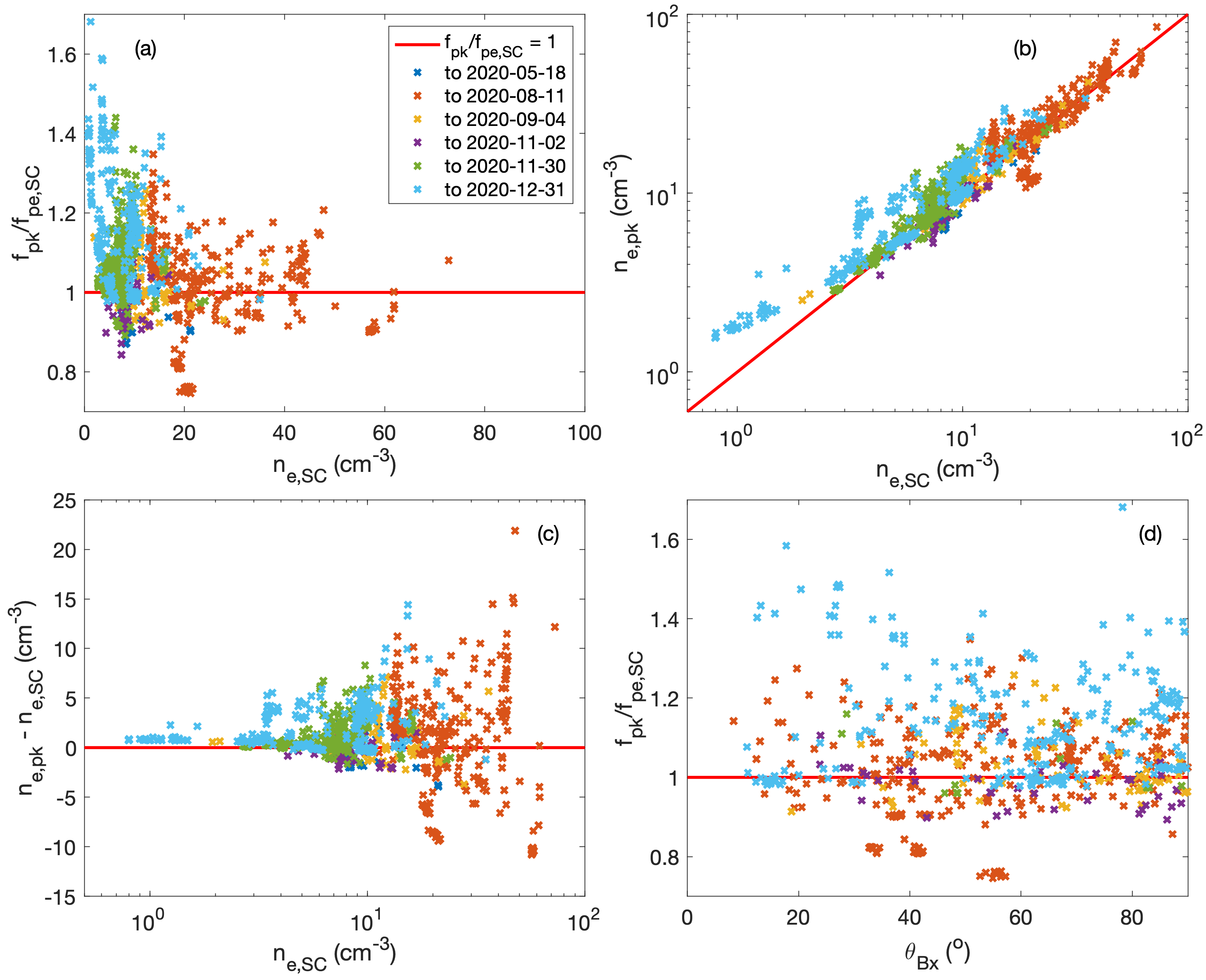}
\caption{Comparison of Langmuir wave frequency $f_{pk}$ with the electron plasma frequency $f_{pe,SC}$ calculated from the spacecraft potential. (a) Scatterplot of $f_{pk}/f_{pe,SC}$ versus $n_{e,SC}$, where each point corresponds to a single TDS snapshot where Langmuir waves are observed. (b) 
Number density $n_{e,pk}$ calculated from Langmuir wave frequency versus $n_{e,SC}$. 
(c) $n_{e,pk} - n_{e,SC}$ versus $n_{e,SC}$. 
(d) $f_{pk}/f_{pe,SC}$ 
versus $\theta_{Bx}$. The different colored symbols correspond to the different time intervals indicated in the legend in panel (a). }
\label{Ldensity}
\end{center}
\end{figure*}

We now investigate statistically the agreement between $n_e$ calculated from the spacecraft potential $n_{e,SC}$ and $f_{pk}$ of Langmuir waves. We search for Langmuir waves when both TDS data and spacecraft potential data are available over 2020. 
To search for and identify Langmuir waves we use the following criteria: (1) The frequency of peak Langmuir wave power $f_{pk}$ lies within $0.5 f_{pe,SC} < f_{pk} < 1.5 f_{pe,SC}$ (we have checked that this range is sufficient to capture the observed Langmuir waves), where $f_{pe,SC}$ is the electron plasma frequency estimated from the spacecraft potential. (2) The peak wave power is two orders above the background power in the frequency range surrounding the Langmuir waves. We inspected each of the snapshots to remove false-positives due to artificial spectral peaks (spacecraft interferences). Over 2020 we identify 961 Langmuir wave events, with 216 waveforms identified in June 2020. 
The statistical results are shown in Figure \ref{Ldensity}. 
In Figure \ref{Ldensity}a we plot $f_{pk}/f_{pe,SC}$ calculated from the spacecraft potential as a function of $n_{e,SC}$. Each point corresponds to a TDS snapshot where we identify Langmuir waves. Overall, we find that most of the points are clustered at $f_{pk}/f_{pe,SC} \approx 1$, with a median value of $1.05$. Throughout 2020 the bias currents to the probes were changed multiple times, meaning different calibrations are required to calculate $n_{e,SC}$. The 
points are color coded by the times when different calibrations were used (the times are given in the legend of Figure \ref{Ldensity}a). In general, the results are similar for each calibration interval, with median values of $0.98$, $1.03$, $1.02$, $0.98$, $1.06$, and $1.13$ for the six time intervals (ordered sequentially in time). For the December interval $f_{pk}/f_{pe,SC}$ typically exceeds $1$ and has the highest median value. We note that some of the clumps of points with $f_{pk}/f_{pe,SC} < 1$ were observed in July 2020 (red crosses), when the bias currents to the probes was non-ideal resulting in a more uncertain $n_{e,SC}$. 
Figure \ref{Ldensity}b shows the density estimated from the Langmuir wave frequency 
$n_{e,pk}$ versus $n_{e,SC}$. We find that there is good agreement between $n_{e,pk}$ and $n_{e,SC}$, confirming that the spacecraft potential is reliable as a density probe. The only exception is at low $n_{e,SC}$ where $n_{e,SC}$ is underestimated. These intervals were primarily observed in December when Solar Orbiter was approximately 1 AU from the Sun. This underestimation of $n_{e,SC}$ is likely due to the uncertainty in the calibration of $n_{e,SC}$ due to the difficulty of identifying the quasi-thermal noise spectral peak at low frequencies. 
However, from Figure \ref{Ldensity}c where we plot $n_{e,pk} - n_{e,SC}$ versus $n_{e,SC}$, we see that the absolute difference between between $n_{e,pk}$ and $n_{e,SC}$ is small at low densities, and tends to increase with increasing $n_{e,SC}$.

\begin{figure}[htbp!]
\begin{center}
\includegraphics[width=90mm, height=40mm]{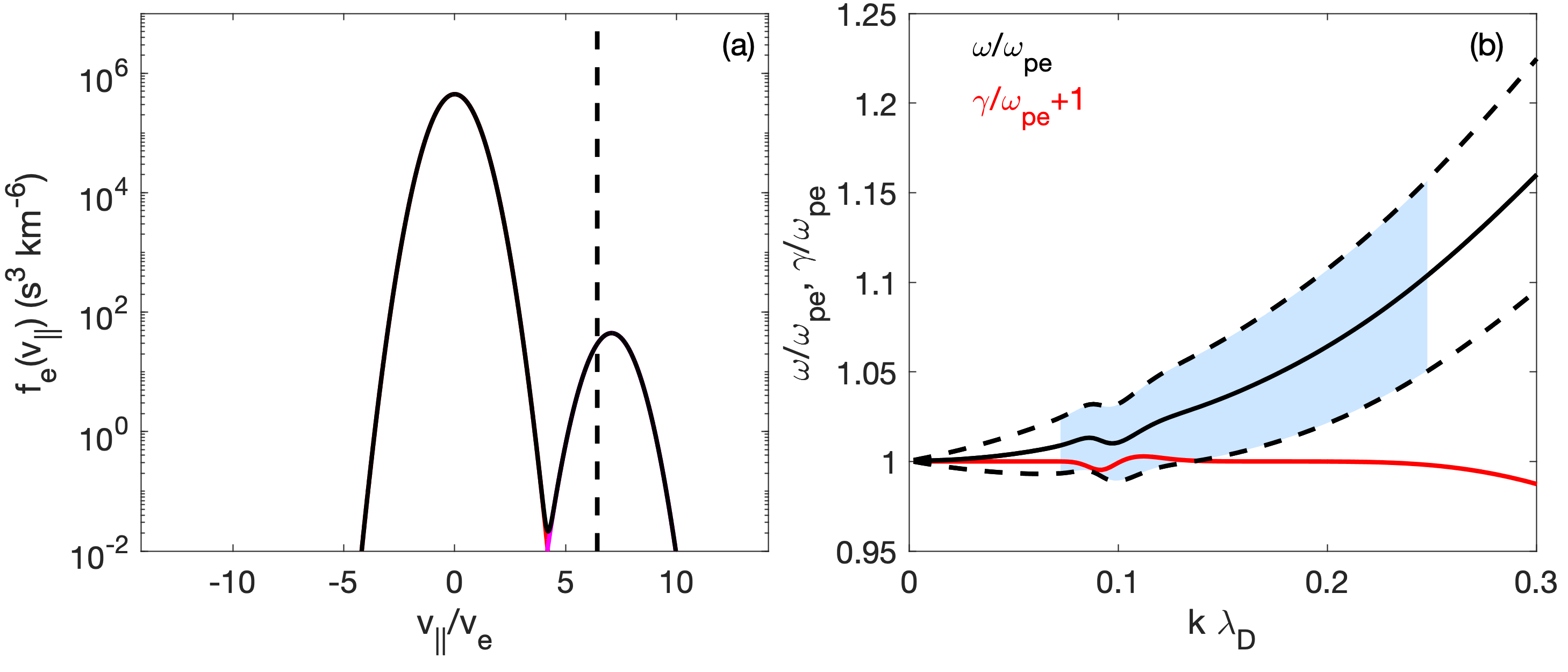}
\caption{Example of the dispersion relation of a Langmuir waves driven by the bump-on-tail instability using equation (\ref{eseq}). (a) The distribution function used to calculate the dispersion relation (black line). The vertical dashed black line shows the $v_{ph}$ of the wave correspond to maximum growth. (b) Dispersion relation of Langmuir waves (black line) and associated growth rate (red line) shifted up in frequency by $\omega_{pe}$. The 
upper and lower dashed lines show the Doppler-shifted dispersion dispersion relation for ${\bf k}$ aligned and anti-aligned with s solar wind flow of $350$~km~s$^{-1}$. The blue shaded region shows the expected range of $k$ and frequencies for Langmuir waves. }
\label{Ldisp} 
\end{center}
\end{figure}

We now investigate whether the observed scatter in $f_{pk}/f_{pe,SC}$ can be attributed to $f_{pk}$ differing from the local electron plasma frequency or is due to the uncertainty in $n_{e,SC}$. The linear dispersion relation of 
Langmuir waves in the spacecraft frame is
\begin{equation}
\omega = \omega_{pe} + \frac{3 v_e^2 k^2}{4 \omega_{pe}} + k V_{sw} \cos{\theta_{Bx}},
\label{Ldisprel}
\end{equation}
where $v_e = \sqrt{2 k_B T_e/m_e}$ is the electron thermal speed and $\theta_{Bx}$ is assumed to be the angle between the solar wind flow and the wave vector ${\bf k}$. The final term in equation (\ref{Ldisprel}) is the Doppler shift due to the plasma flow past the spacecraft. There are two effects that can cause $f_{pk}$ to differ from $f_{pe}$, namely, the increase in frequency due to the thermal correction, and Doppler shift. Note 
that Doppler shift can be both positive and negative. These effects are illustrated in Figure \ref{Ldisp}, where we plot the dispersion relation of Langmuir waves driven by a weak electron beam. For simplicity we use a single Maxwellian for the background electron distribution, neglecting the halo and Strahl contributions. Figure \ref{Ldisp}a shows an electron distribution consisting of a core population ($n_e = 30$~cm$^{-3}$ and $T_e = 15$~eV) and a weak beam ($n_b = 3 \times 10^{-3}$~cm$^{-3}$, $T_b = 15$~eV and $V_b = 7.1 v_e$), which is unstable to Langmuir waves. Figure \ref{Ldisp}b show the dispersion resulting dispersion relation (solid black line) for $\theta_{Bx} = 90^{\circ}$ (equivalent to the dispersion relation in the plasma rest frame). The fluctuation in $\omega$ around $k \lambda_D = 0.1$ is due to the electron beam. The red line indicates the growth and damping rate $\gamma$. For these parameters $\gamma > 0$ peaks for $k \lambda$ just above 0.1 (the predicted wave number is $k \lambda_D = \omega_{pe} \lambda_D/v_b = 0.1 \lambda_D$ to simplest approximation). 
The upper and lower black dashed lines show the Doppler-shifted dispersion relations for outwardly directed ${\bf k}$ and inwardly directed ${\bf k}$, respectively, where $V_{sw} = 350$~km~s$^{-1}$ is used. Figure \ref{Ldisprel}b shows that there can be a significant change in the Langmuir wave frequency from $f_{pe}$, if $k \lambda_D$ becomes large enough. We note that it is possible for $f_{pk}$ to be slightly less than $f_{pe}$ if ${\bf k}$ is directed Sunward. At lower $k \lambda_D$ the deviation in wave frequency from $f_{pe}$ is primarily due to Doppler shift, while at larger $k \lambda_D$ the thermal correction becomes more prominent. This is because the Doppler shift increases linearly with $k$, while the thermal correction is quadratic in $k$. 

We can provide a rough estimate of the wave numbers likely to be seen in the solar wind. From observations we find that the
majority of Langmuir waves are approximately field-aligned $E_{\parallel} \gg E_{\perp}$. From previous observations of type III source regions it was found that $E_{\parallel} \gg E_{\perp}$ was observed for $v_b/c \gtrsim 0.08$, whereas for $v_b/c \lesssim 0.08$ Langmuir waves with strong perpendicular electric fields would also develop \cite[]{malaspina2011,graham2013,graham2014}. This suggests that for most cases in the solar wind $v_b/c \lesssim 0.08$, which corresponds to $k \lambda_D \gtrsim 0.07$. From Figure \ref{Ldisp}b we find that Landau damping starts to become significant for $k \lambda_D \gtrsim 0.25$. Similarly, to excite Langmuir waves electron beams are expected to satisfy $V_b \gtrsim 3 v_e$, corresponding to $k \lambda_D \lesssim 0.24$. This suggests the the largest probable wave number is $k \lambda_D \sim 0.25$. This interval and the associated frequency range due to Doppler shift is indicated by the blue shaded region in Figure \ref{Ldisp}b. 
Based on this range of $k \lambda_D$ we find that the observed range of frequencies is $0.98 \lesssim \omega/\omega_{pe} \lesssim 1.16$, although this frequency range will increase as $V_{sw}$ increases. We find that the overall median $f_{pk}/f_{pe}$ is near the center of this predicted frequency range, and that $\sim 60 \%$ of the events in Figure \ref{Ldensity} lie within this frequency range. 

From Figure \ref{Ldisp}b we see that the Doppler shifted dispersion relations have a wider range of frequencies as a function of $k$ compared with dispersion relation without Doppler shift or $\theta_{Bx} = 90^{\circ}$. Therefore, if the scatter in $f_{pk}/f_{pe,SC}$ is due to the variability of the Langmuir wave frequency, and hence $k$, rather than uncertainties in $f_{pe,SC}$, we expect to see more in $f_{pk}/f_{pe,SC}$ as $\theta_{Bx}$ decreases. In Figure \ref{Ldensity}d we plot $f_{pk}/f_{pe,SC}$ versus $\theta_{Bx}$. We find little dependence on the scatter in $f_{pk}/f_{pe,SC}$ as a function $\theta_{Bx}$. This suggests that the uncertainty in $n_{e,SC}$ is too large to accurately resolve the difference in $f_{pk}$ and $f_{pe}$. This is perhaps unsurprising because very precise estimates of $n_{e,SC}$ are required and 
the statistical data incorporates multiple bias changes as the distance of Solar Orbiter from the Sun changes over its first orbit. However, we propose one situation below where it might be possible to quantify the difference between Langmuir wave frequency and $f_{pe}$, and hence provide an estimate of $k$. 

\begin{figure*}[htbp!]
\begin{center}
\includegraphics[width=170mm, height=100mm]{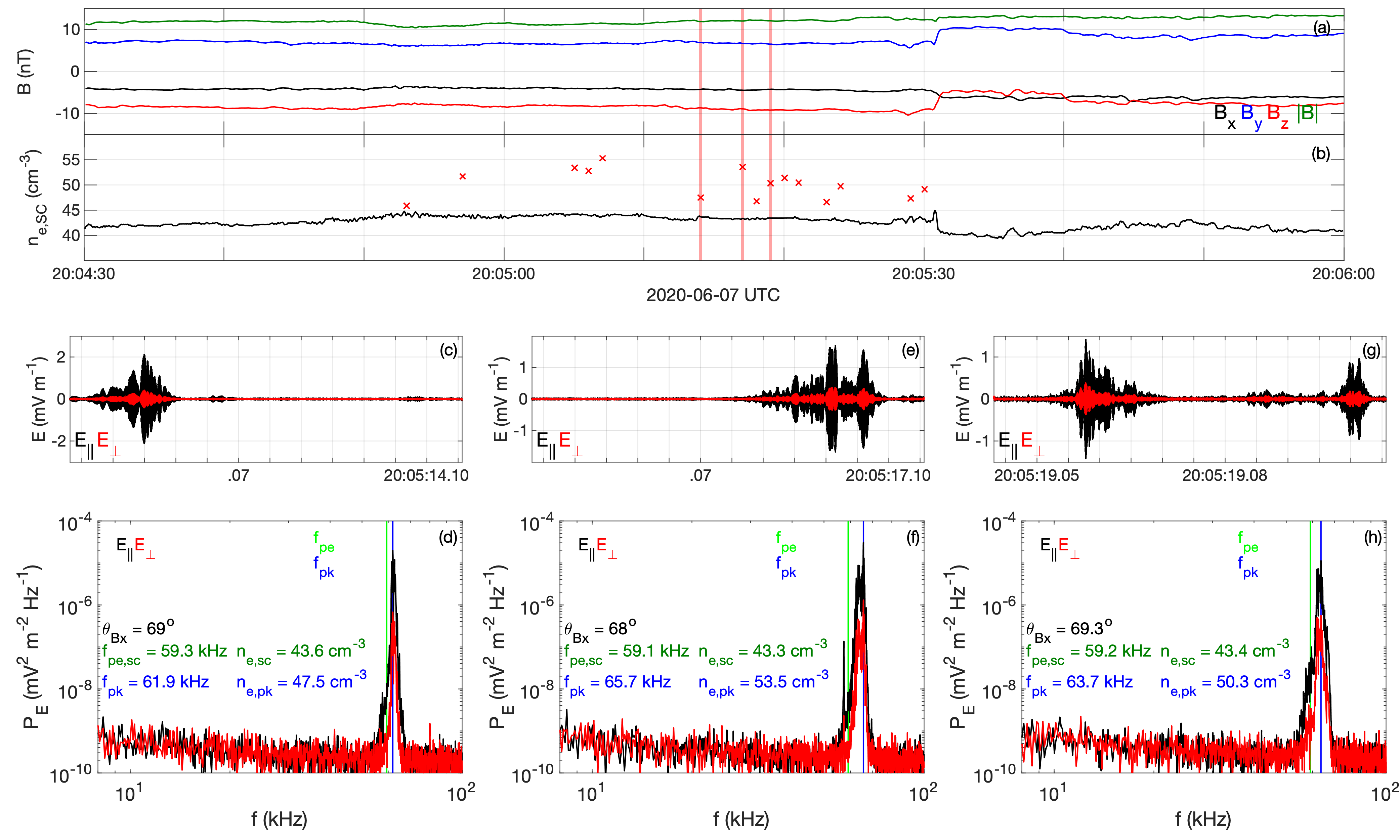}
\caption{Langmuir waves observed near a solar wind current sheet on 2020 June 07. A series of Langmuir waves are observed with variable frequencies above the electron plasma frequency predicted from $n_{e,SC}$. (a) ${\bf B}$. (b) $n_{e,SC}$. 
The red crosses indicates the TDS snapshots with Langmuir waves and the number density estimated from $f_{pk}$. The three red shaded regions correspond to the three Langmuir wave examples in panels (c)--(h). 
Panels (c), (e), (g) show three Langmuir waveforms in field-aligned coordinates $E_{\parallel}$ (black) and $E_{\perp}$ (red). Panels (d), (f), (g) show the power spectra of $E_{\parallel}$ (black) and $E_{\perp}$ (red) associated with the Langmuir waveforms in panels (c), (e), and (g), respectively. }
\label{LegCS}
\end{center}
\end{figure*}

The TDS receiver is triggered by wave activity so in some cases a series of Langmuir wave snapshots can be captured in rapid succession, where the solar wind conditions are approximately constant. Figure \ref{LegCS} shows a solar wind interval observed on 2020 June 07 where we see a series of Langmuir wave snapshots. Figures \ref{LegCS}a and \ref{LegCS}b show ${\bf B}$ and $n_{e,SC}$ over a 90 second interval. The red crosses in Figure \ref{LegCS}b show the times of the Langmuir wave snapshots and $n_{e,pk}$ calculated from the peak Langmuir frequency. In this interval we find $15$ Langmuir wave snapshots over a 40 second interval. 
The Langmuir waves are observed just prior to a narrow low-shear current sheet. When the waves are observed the solar wind conditions are approximately constant meaning that changes in $\theta_{Bx}$ and $f_{pe}$ are negligible. In contrast, $n_{e,pk}$ and $f_{pk}$ change significantly, which suggests that $k$ is changing, based on equation (\ref{Ldisprel}). In particular $n_{e,pk}$ varies from $45$~cm$^{-3}$ to $55$~cm$^{-3}$ (corresponding to $61 \, \mathrm{kHz} \lesssim f_{pk} \lesssim 67 \, \mathrm{kHz}$), while $n_{e,SC}$ remains relatively constant at $43 - 44$~cm$^{-3}$. 

In Figures \ref{LegCS}c--\ref{LegCS}g we show the waveforms of three Langmuir wave snapshots and their associated power spectra, corresponding to the three red-shaded intervals in Figures \ref{LegCS}a and \ref{LegCS}b. In all three cases the Langmuir waves are quite localized and $E_{\parallel} \gg E_{\perp}$, similar to the event in Figures \ref{Leg}d--\ref{Leg}f. In each case the spectral peaks are quite narrow, indicating Langmuir waves rather than beam-mode waves, and $f_{pk}$ occurs above $f_{pe,SC}$. However, we find that $f_{pk}$ differs in each case, while $f_{pe,SC}$ and $\theta_{Bx}$ remain constant. This suggests that the $k$ is changing significantly between snapshots. If we assume a constant $n_e = 43.5$~cm$^{-3}$ based on the average $n_{e,SC}$ we can estimate $k$ of the Langmuir waves. From equation (\ref{Ldisprel}) we obtain
\begin{equation}
k = \frac{2 \sqrt{\omega_{pe}^2 V_{sw}^2 \cos^2{\theta_{Bx}} + 3 \omega_{pe} v_e^2 (\omega - \omega_{pe})} - 2 \omega_{pe} V_{sw} \cos{\theta_{Bx}}}{3 v_e^2}.
\label{kval}
\end{equation}
In equation (\ref{kval}) we have assumed ${\bf k}$ is directed outward from the Sun. We expect this to generally be the case, as electron beams exciting the Langmuir waves should originate Sunward of the spacecraft. Although there may be some cases where ${\bf k}$ is directed Sunward, such as backscattered Langmuir waves produced by three-wave electrostatic decay \cite[e.g.][]{cairns1987a}. Over the time interval in Figure \ref{LegCS} we do not have any particle data so we use the nominal $T_e = 15$~eV and $V_{sw} = 350$~km~s$^{-1}$. Using these parameters and equation (\ref{kval}) we calculate $k \lambda_D = 0.15$, $0.24$, and $0.20$ for the Langmuir waves in Figures \ref{LegCS}c and \ref{LegCS}d, Figures \ref{LegCS}e  and \ref{LegCS}f, and Figures \ref{LegCS}g and \ref{LegCS}h, respectively. These values all lie within the range of expected $k \lambda_D$ shown in Figure \ref{Ldisp}b, which supports this method of estimating $k$ being reasonable and $n_{e,SC}$ being reliable. For the Langmuir waves in this interval we estimate the range of $k$ to be $0.11 \lesssim k \lambda_D \lesssim 0.27$. These values are reasonable and lie within the range of $k$ where $E_{\parallel} \gg E_{\perp}$ is expected \cite[]{graham2014}, which is consistent with observations. The changes in $k$ could result from either changes in electron beam speeds or low-amplitude density fluctuations \cite[]{smith1979,robinson1992,voshchepynets2015}. Particle data is required for accurate estimations of $V_{sw}$ and $v_e$ and more reliable estimations of $k$. However, these results show that the variability of $f_{pk}$ with respect to $f_{pe,SC}$ can in part result from the variability of $k$ of Langmuir waves.

In summary, we find Langmuir waves through the first orbit of Solar Orbiter in the solar wind, which are not associated with type II or type III source regions. We have compared the Langmuir wave frequencies with the the electron plasma frequency calculated from the spacecraft potential. The results show that the spacecraft potential is a reliable probe of the background electron plasma density. The variability of the Langmuir wave frequency with respect to the estimated electron plasma frequency can in part be explained by the variability of the wave number of Langmuir waves. 

\section{Solar wind currents and waves} \label{swcurrents}
In this section we estimate the currents in the solar wind and compare the occurrence of strong currents with ion-acoustic and Langmuir waves. We use two methods to identify strong currents and current structures in the solar wind: 

(1) We estimate current densities from the changes in ${\bf B}$ by assuming the current structures are frozen in to the solar wind flow. If we assume the current structures are moving with the solar wind flow we can estimate the current densities in the y-z plane using 
\begin{equation}
J_{y} = - \frac{1}{\mu_0} \frac{\Delta B_z}{V_{sw} \Delta t}, J_{z} = \frac{1}{\mu_0} \frac{\Delta B_y}{V_{sw} \Delta t},
\label{Jcalc}
\end{equation}
Since the particle data is only intermittently available over June 2020 we simply assume $V_{sw} = -350$~km~s$^{-1}$ in the x-direction, which is close to the median value calculated when ion moments are available. 
We note that this estimate of ${\bf J}$ is most reliable when the normal to the current structure is in the x-direction, as this method assumes $J_x = 0$. In cases where the normal is highly oblique to the x-direction we expect ${\bf J}$ to be underestimated. 

(2) We use the Partial Variance of Increments (PVI) method to identify strong discontinuities in the solar wind. 
For single spacecraft measurements the PVI value is given by \cite[]{greco2008}
\begin{equation}
PVI = \frac{|\Delta {\bf B}(t,\tau)|}{\sqrt{\langle|{\bf B}(t,\tau)|^2\rangle}}, 
\label{PVIeq}
\end{equation}
where $\Delta {\bf B}(t,\tau) = {\bf B}(t + \tau) - {\bf B}(t)$, $\tau$ is the separation in time, and $\langle ... \rangle$ indicates the average. Here, we calculate PVI over the entire June 2020 interval. The PVI value is increased by both changes in the magnitude of ${\bf B}$ and rotational changes \cite[]{greco2018}. 

\begin{figure}[htbp!]
\begin{center}
\includegraphics[width=90mm, height=40mm]{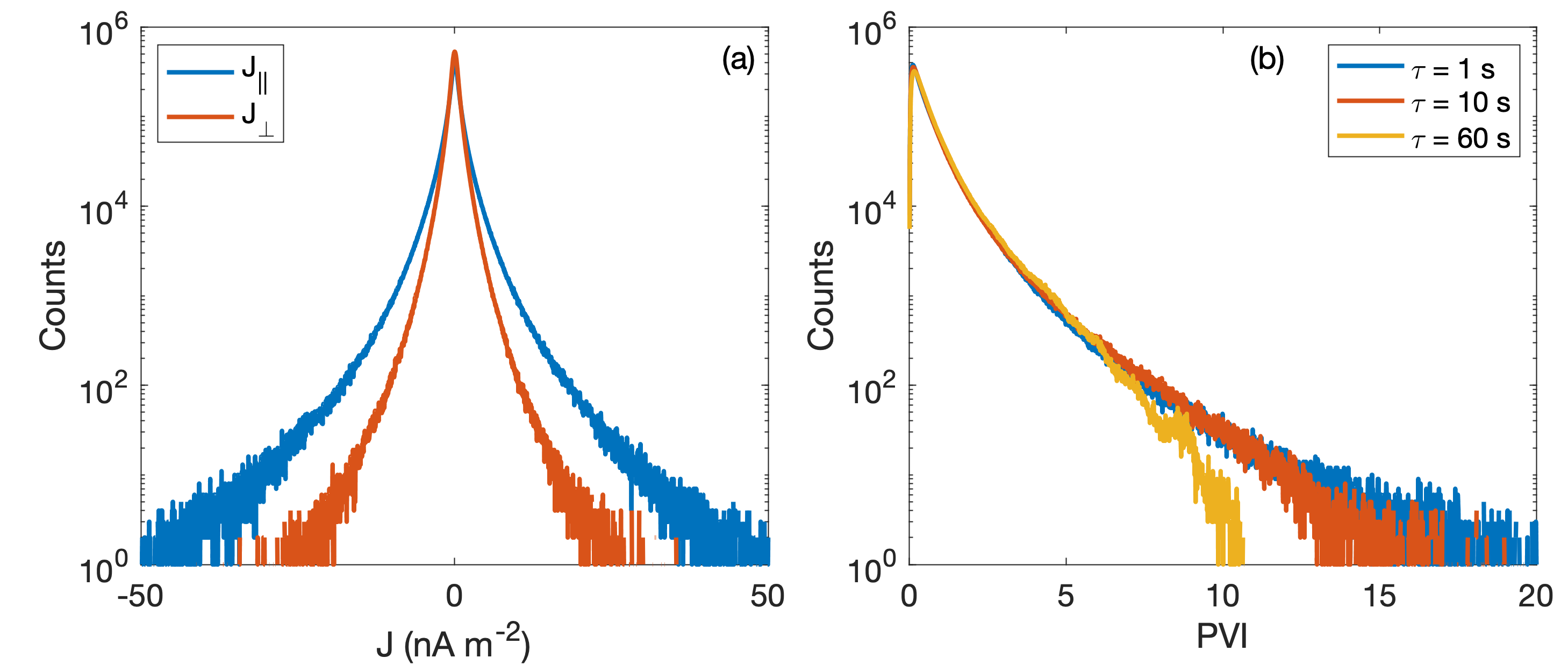}
\caption{Statistics of ${\bf J}$ and PVI over June 2020. (a) Histogram of $J_{\parallel}$ (blue) and $J_{\perp}$ (red). (b) Histogram of PVI using $\tau = 1$~s (blue), $10$~s (red), $60$~s (gold). }
\label{JPVI}
\end{center}
\end{figure}

To provide an overview of the currents observed over June 2020 we plot the histograms of ${\bf J}$ and PVI values over the entire month in Figure \ref{JPVI}. Figure \ref{JPVI}a shows the histograms of $J_{\parallel}$ and $J_{\perp}$ 
(where we have used the same coordinate transformation as for ${\bf E}$). As expected we find that the distributions peak at $J = 0$ and are non-Gaussian. We find that the maximum values of $J_{\parallel}$ are $\sim 50$~nA~m$^{-2}$. If we compare with the threshold currents required for instability based on Figure \ref{minst}, we find that only the ion beam driven case has a threshold $J$ comparable to the maximum observed $J_{\parallel}$. For the simple electron-ion streaming instability we find that the maximum observed $J_{\parallel}$ is almost two orders of magnitude smaller than the threshold $J$ required for instability. We note that even though $J$ can be underestimated, for example, when the solar wind is slow or when current sheet normals are approximately perpendicular to the solar wind flow. However, it is highly implausible that $J_{\parallel}$ is statistically underestimated by such a degree to make the electron-ion streaming instability possible. 
We conclude that the simple electron-ion streaming instability is unlikely to account for any of the observed ion-acoustic waves. For the electron-electron-ion streaming instability we find that the observed $J_{\parallel}$ is well below the threshold based on Figure \ref{minst}. However, this instability does not have a definite threshold $J$ because of the number of free parameters associated with the core electrons and electron beam. Additionally, it is possible that the instability can occur for $J = 0$, when the core and beam electrons have bulk velocities in opposite directions with respect to the ion population. Therefore, the electron-electron-ion instability remains a possible source of ion-acoustic waves. 

In Figure \ref{JPVI}b we plot the histograms of the PVI values for $\tau = 1$~s, $\tau = 10$~s, and $\tau = 60$~s. For reference, for the median solar wind conditions the ion inertial length is $d_i \approx 40$~km, which translates to a timescale of $\sim 0.1$~s. We use $\tau = 1$~s to identify ion-scale current structures, while 
$\tau = 10$~s and $\tau = 60$~s will identify larger MHD scale current structures. The histograms for $\tau = 1$~s and $\tau = 10$~s exhibit similar histograms, with largest values of PVI being $\approx 20$. For $\tau = 60$~s the histogram is similar at low values, while the PVI values peaks around $10$. Below we consider the current structures to be strong when the PVI value exceeds $5$. For each $\tau$ we find that $\approx 0.4 \%$ of points exceed 5. However, the number of points with PVI $> 5$ associated with a single structure increases as $\tau$ increases. As a result the number of distinct current structures for $\tau = 1$~s is over an order of magnitude higher than the number identified for $\tau = 60$~s. 

\begin{figure}[htbp!]
\begin{center}
\includegraphics[width=90mm, height=90mm]{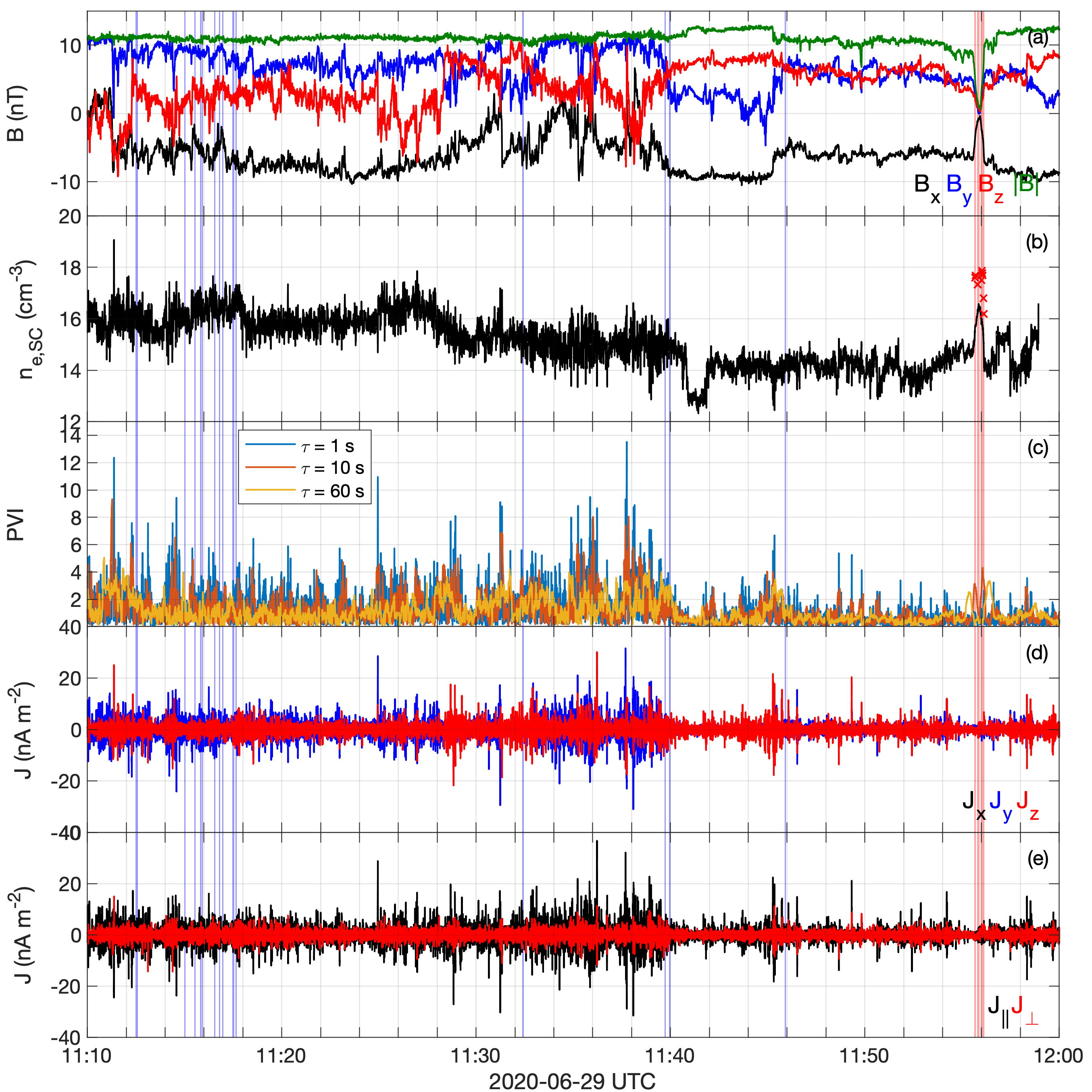}
\caption{A turbulent solar wind interval with a clear magnetic hole observed by Solar Orbiter on 2020 June 29. (a) ${\bf B}$. (b) $n_{e,SC}$. The red crosses are $n_{e,pk}$ estimated from Langmuir waves. 
(c) PVI values for $\tau = 1$~s (blue), $10$~s (red), and $60$~s (gold). 
(d) and (e) ${\bf J}$ in SRF and field-aligned coordinates, respectively. The blue-shaded and red-shaded regions indicate times when ion-acoustic and Langmuir wave snapshots are observed by TDS, respectively.}
\label{CSeg1}
\end{center}
\end{figure}

To further investigate where the ion-acoustic waves and Langmuir waves are observed in relation to current structures, we show some specific solar wind intervals. In Figure \ref{CSeg1} we plot a 50 minute solar wind interval on 2020 June 29. We plot ${\bf B}$, $n_{e,SC}$, PVI values, and ${\bf J}$, and mark the times TDS observed ion-acoustic waves (blue-shaded regions) and Langmuir waves (red-shaded regions). Figure \ref{CSeg1}a shows that the interval is quite turbulent. While $|{\bf B}|$ remains relatively constant, current sheets are observed, in addition to fluctuations in ${\bf B}$. The fluctuations in ${\bf B}$ and current sheets are primarily observed from 11:10 to 11:40 UT. We observe enhanced fluctuations in $n_{e,SC}$ in association with the fluctuations in ${\bf B}$. 

Throughout this interval we observe 18 ion-acoustic wave snapshots and 9 Langmuir wave snapshots from TDS. Most of the ion-acoustic waves are observed during the first half of the interval when the turbulent fluctuations are larger. By comparing the times of the ion-acoustic wave snapshots with when we see the largest PVI values and ${\bf J}$, we see that the ion-acoustic waves occur when fluctuations in ${\bf J}$ are observed. However, the snapshots do not tend to occur when the PVI values or ${\bf J}$ peaks. This might suggest that the ion-acoustic waves are occurring in turbulent solar wind regions rather than being generated at the current sheets themselves. 
In Figure \ref{CSeg1} we observe all the Langmuir wave snapshots within the magnetic hole seen at 11:56 UT. The magnetic hole is observed over an approximately 40 second period. We find other magnetic holes with simultaneous Langmuir wave observations in the June 2020 period (not shown). The tendency of solar wind Langmuir waves to occur within magnetic holes was observed at 1 AU by \cite{braind2010}, and this tendency is also present closer to the Sun at 0.5 AU. Figure \ref{CSeg1}b also shows $n_{e,pk}$ associated with the Langmuir waves. Like the event in Figure \ref{Ldisp} we find that $n_{e,pk}$ varies, indicating that $f_{pk}$ changes between snapshots. However, in this case the density varies across the magnetic hole to maintain pressure balance, so it is more difficult to determine if $k$ is changing for the Langmuir waves. 

\begin{figure}[htbp!]
\begin{center}
\includegraphics[width=90mm, height=90mm]{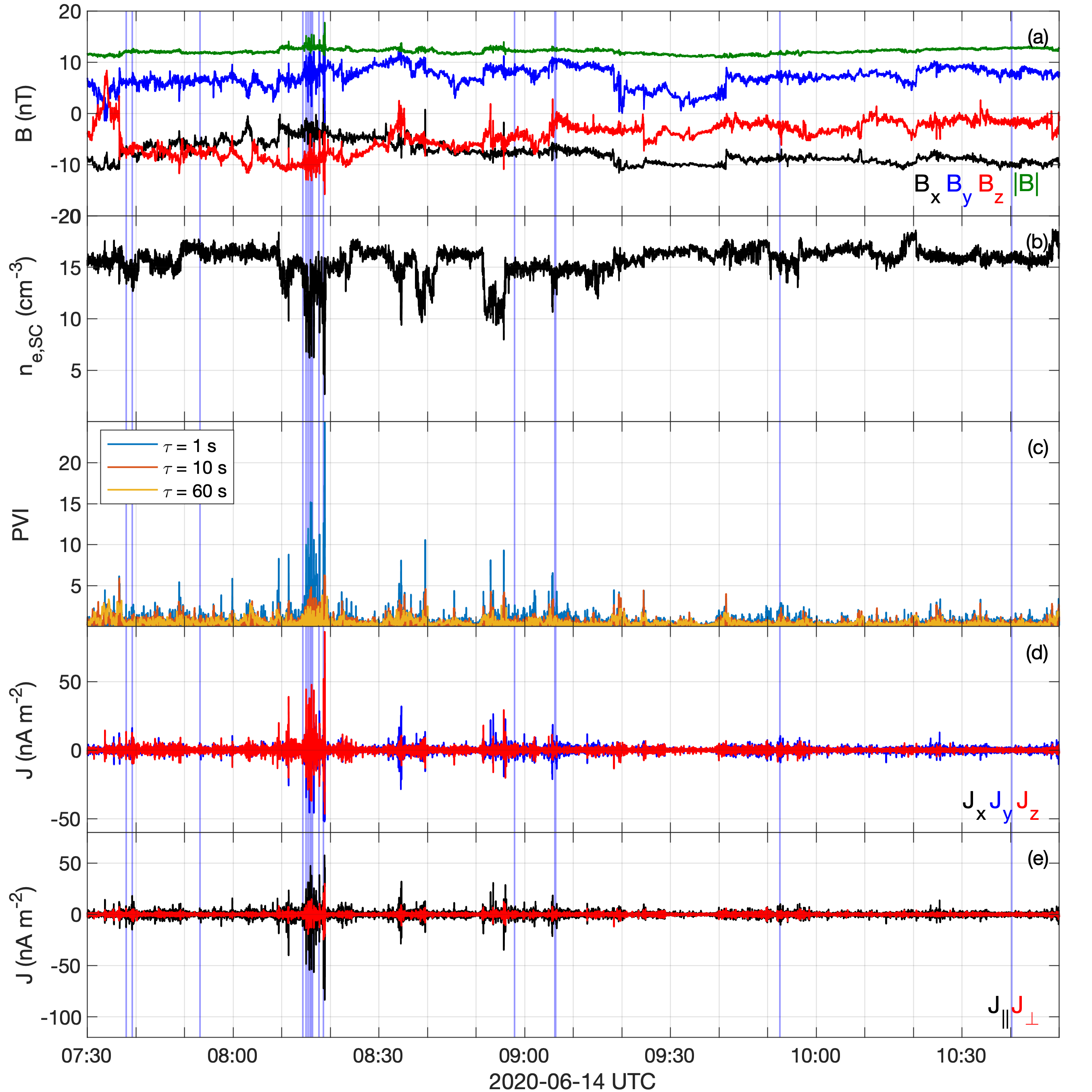}
\caption{A solar wind interval by Solar Orbiter on 2020 June 14. (a) ${\bf B}$. (b) $n_{e,SC}$. 
(c) PVI values for $\tau = 1$~s (blue), $10$~s (red), and $60$~s (gold). 
(d) and (e) ${\bf J}$ in SRF and field-aligned coordinates, respectively. The blue-shaded regions indicate times when ion-acoustic wave snapshots are observed by TDS.}
\label{CSeg2}
\end{center}
\end{figure}

As a second example of when ion-acoustic waves are observed, Figure \ref{CSeg2} shows a 3-hour solar wind interval from 2020 June 14. The interval in Figure \ref{CSeg2} is quieter than in Figure \ref{CSeg1}. However, between 08:10 and 08:20 UT we observed very strong currents, which are amongst the strongest seen over June 2020 with a peak $J$ close to $100$~nA~m$^{-2}$. These currents are primarily aligned with ${\bf B}$, although perpendicular currents are also present. 
The currents are associated with rapid rotations in ${\bf B}$. At these structures $|{\bf B}|$ also increases, which is responsible for $J_{\perp}$. When $|{\bf B}|$ increases there are sharp decreases in $n_{e,SC}$, meaning strong density and pressure gradients are also present. At this time the largest PVI values are found, which peak at $\approx 20$ for $\tau = 1$~s. For  $\tau = 10$~s and $60$~s we find significantly smaller PVI values, indicating that the current structures have ion spatial scales. 

We find that most of ion-acoustic wave snapshots occur within this region of strong currents. We identify 19 ion-acoustic wave snapshots over the entire interval, with 11 of the snapshots being found in the region of strong currents. The remaining snapshots we observed when $J$ was substantially smaller and spread out over the entire interval. In some cases these waves are observed around smaller enhancements in $J$, while at other times ion-acoustic waves were observed when $J$ was negligible. 

From Figures \ref{CSeg1} and \ref{CSeg2} we see that in some instances there is a strong correlation of ion-acoustic waves and enhanced currents, which might suggest that large $J$ is crucial to the generation of ion-acoustic waves. At other times we see that the detection of waves occurs in regions of strong solar wind turbulence but the ion-acoustic waves are not strongly correlated with specific current structures. Finally, we sometimes see ion-acoustic waves in quiet regions of the solar wind, which do not appear to be associated with any currents. We now consider more statistically the relation between ion-acoustic wave observations and solar wind current structures. 

\begin{figure}[htbp!]
\begin{center}
\includegraphics[width=90mm, height=80mm]{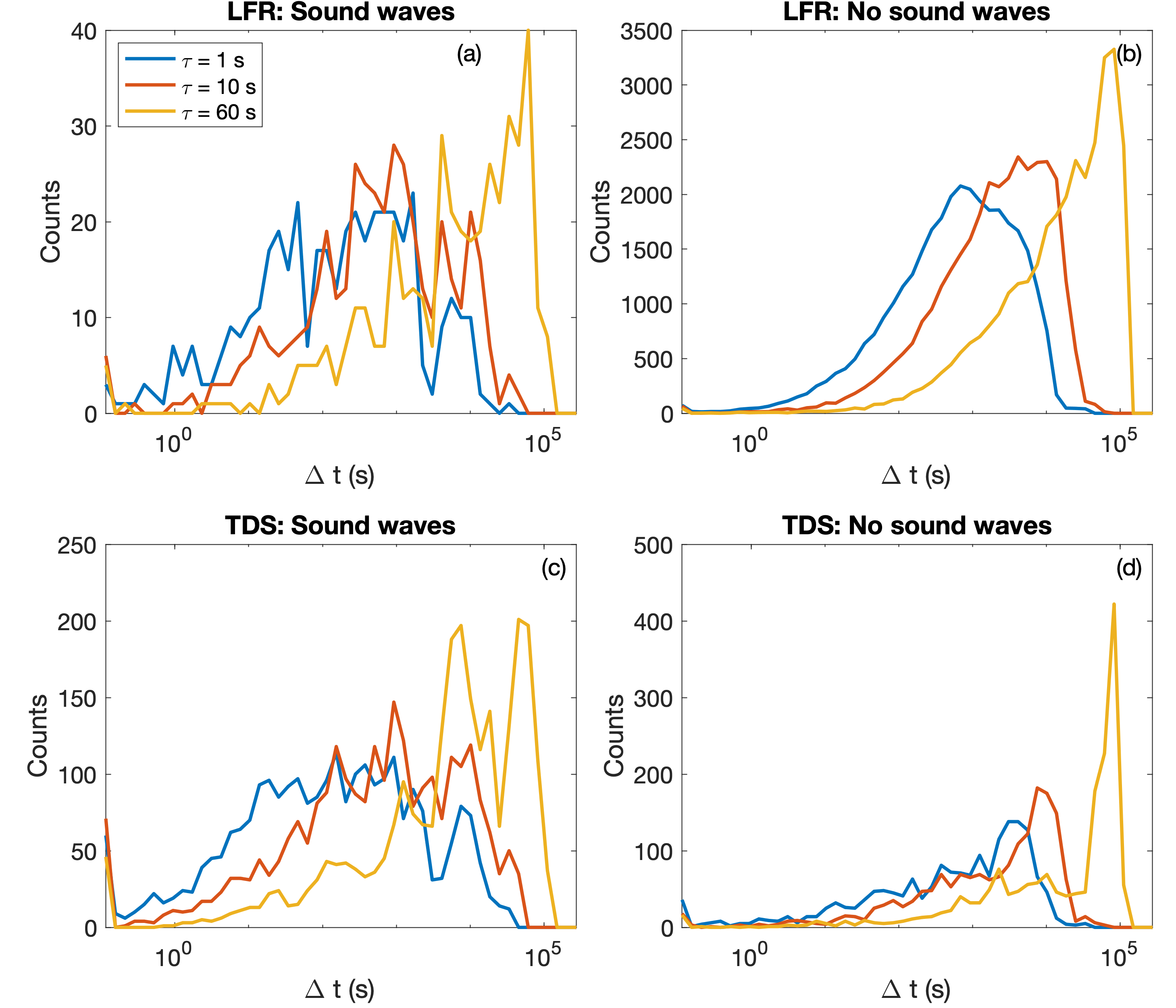}
\caption{Histograms of the time between observations of ion-acoustic waves $\Delta t$ from LFR and TDS and the nearest strong current structure identified by PVI values greater than 5 for $\tau = 1$~s (blue), $10$~s (red), and 
$60$~s (gold). (a) Histograms of $\Delta t$ for ion-acoustic waves observed by LFR. (b) Histograms of $\Delta t$ for ion-acoustic waves observed by LFR. (b) Histograms of $\Delta t$ for LFR snapshots where we do not identify ion-acoustic waves. (c) Histograms of $\Delta t$ for ion-acoustic waves observed by TDS. (c) Histograms of $\Delta t$ for ion-acoustic waves observed by LFR. (d) Histograms of $\Delta t$ for TDS snapshots where we do not identify ion-acoustic waves. The points at $\Delta t = 10^{-1}$~s are snapshots at or within current structures. }
\label{CSstats}
\end{center}
\end{figure}

We first consider the time between ion-acoustic wave observations and the nearest strong current structure, defined as having a maximum PVI value exceeding 5. 
We calculate the time $\Delta t$ between the snapshot and the nearest strong current structure. We do this for all LFR and TDS snapshots over June 2020, for both snapshots where ion-acoustic waves are observed and snapshots where we did not observe ion-acoustic waves. The results are shown in Figure \ref{CSstats}, where we plot the histograms of $\Delta t$ for $\tau = 1$~s, $10$~s, and $60$~s. The counts at the smallest $10^{-1}$~s correspond to snapshots observed at or within strong current structures. In Figures \ref{CSstats}a and 
\ref{CSstats}b we plot the histograms of $\Delta t$ for LFR snapshots with ion-acoustic waves and LFR snapshots without ion-acoustic waves, respectively. Since the LFR snapshots are taken at regular times, the histograms in Figure \ref{CSstats}b provide distributions of the probable $\Delta t$ regardless of whether ion-acoustic waves are observed or not (note that about $~1 \%$ of all LFR snapshots contained ion-acoustic waves). In both Figures \ref{CSstats}a and \ref{CSstats}b we see that $\Delta t$ increases as $\tau$ increases. This simply corresponds to smaller-scale current structures occurring more regularly than larger scale structures. This can be seen in Figures \ref{CSeg1} and \ref{CSeg2}. We find that when ion-acoustic waves are present the distribution of $\Delta t$ is shifted to smaller values. However, only a small fraction of the ion-acoustic waves are observed at the strong current structures themselves. Most ion-acoustic waves are observed several minutes from the nearest ion-scale current structures ($\tau = 1$~s). For ion-acoustic wave snapshots the median $\Delta t$ are $3$~minutes, $11$~minutes, and $2$~hours for $\tau = 1$~s, $10$~s, and $60$~s, respectively. At times when no ion-acoustic waves are observed in the LFR snapshots, corresponding to regularly sampled times in the solar wind, the median $\Delta t$ are $11$~minutes, $40$~minutes, and $6$~hours for $\tau = 1$~s, $10$~s, and $60$~s, respectively. Therefore, $\Delta t$ are reduced when ion-acoustic waves are present.

In Figures \ref{CSstats}c and \ref{CSstats}d we plot the histograms $\Delta t$ for TDS snapshots with and without ion-acoustic waves. In contrast to LFR, comparable numbers of snapshots with and without ion-acoustic waves are recorded. In Figure \ref{CSstats}c we find that the histograms of $\Delta t$ are similar to those from LFR, and like the results in Figure \ref{CSstats}a we find that very few ion-acoustic waves were observed at the current structures themselves. We find that the median $\Delta t$ are  $2$~minutes, $11$~minutes, and $2$~hours for $\tau = 1$~s, $10$~s, and $60$~s, respectively, which agrees with the LFR results. When no ion-acoustic waves are observed by TDS we find significantly larger $\Delta t$, with 
median values of $15$~minutes, $50$~minutes, and $12$~hours for $\tau = 1$~s, $10$~s, and $60$~s, respectively. Overall, we find that ion-acoustic waves are more likely to be observed closer to strong current structures compared to solar wind times without ion-acoustic waves. However, only a small fraction of the observed ion-acoustic waves occur at the strong current structures themselves. 

\begin{figure*}[htbp!]
\begin{center}
\includegraphics[width=150mm, height=90mm]{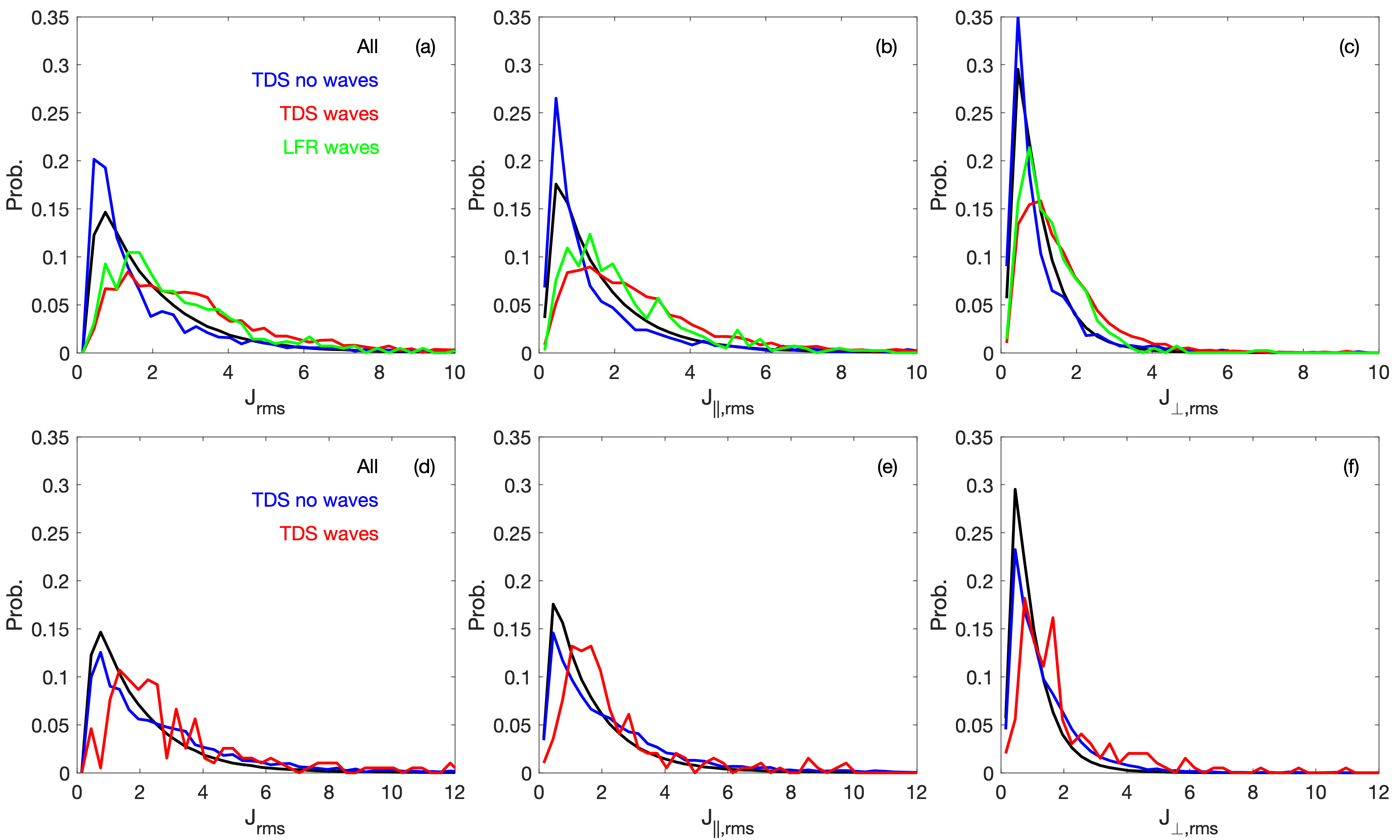}
\caption{Histograms of the root-mean-square currents $J_{rms}$ at the times when ion-acoustic and Langmuir waves are observed. $J_{rms}$ are computed over 10~s intervals with the snapshot time centered in the middle. 
(a)--(c) $J_{rms}$, $J_{\parallel,rms}$, and $J_{\perp,rms}$ for ion-acoustic wave intervals. The black curves are $J_{rms}$ of all June 2020, the blue curves are TDS intervals with no ion-acoustic waves, the red curves are TDS intervals with ion-acoustic waves, and the green curves are LFR intervals with ion-acoustic waves. (d)--(f) $J_{rms}$, $J_{\parallel,rms}$, and $J_{\perp,rms}$ for Langmuir wave intervals. The black curves are $J_{rms}$ of all June 2020, the blue curves are TDS intervals with no Langmuir waves, the red curves are TDS intervals with Langmuir waves.}
\label{CSWstats}
\end{center}
\end{figure*}

Finally, we consider statistically the local currents associated with ion-acoustic waves and compare them to the typical solar wind conditions to see if ion-acoustic waves are correlated with enhanced currents. To quantify the currents around the observed ion-acoustic waves we calculate the root-mean-square current $J_{rms}$ over a 10~s interval, with the snapshot in the middle of the interval. We calculate these $J_{rms}$ for ion-acoustic wave snapshots observed by TDS, and TDS snapshots without ion-acoustic waves. We also calculate $J_{rms}$ over the entire month for comparison. 
Figures \ref{CSWstats}a--\ref{CSWstats}c the histograms of $J_{rms}$, $J_{\parallel,rms}$, and $J_{\perp,rms}$, respectively for all solar wind data (black), TDS snapshots with no ion-acoustic waves (blue), TDS snapshots with ion-acoustic waves (red), and LFR snapshots with ion acoustic waves (green). In each panel we see that when ion-acoustic waves are observed the distributions are shifted to larger $J_{,rms}$, indicating that ion-acoustic waves tend to occur in regions of enhanced current. We find that $J_{\parallel,rms}$, and $J_{\perp,rms}$ are both statistically larger when ion-acoustic waves are present. 
Over the June 2020 we find a median $J_{rms}$ of $1.5$~nA~m$^{-2}$, while when ion-acoustic waves are observed by LFR and TDS the median $J_{rms}$ are $2.2$~nA~m$^{-2}$ and $2.7$~nA~m$^{-2}$. This suggests that ion-acoustic waves are more likely to be seen in the more turbulent solar wind where the typical fluctuations in ion-scale currents are larger.

In Figures \ref{CSWstats}c--\ref{CSWstats}e we perform the same analyses for Langmuir wave snapshots observed by TDS. We find that Langmuir waves in the solar wind are associated enhanced currents, similar to ion-acoustic waves. When Langmuir waves are present we find a median $J_{rms}$ of $2.4$~nA~m$^{-2}$, which is comparable to the values obtained for ion-acoustic waves. 

In summary we have compared the observation of ion-acoustic and Langmuir waves with the local plasma conditions, focusing on the current density and current structures. We find that the largest observed currents are well below the threshold required for the electron-ion streaming instability. More complex streaming instabilities such as ion-beam instability or electron-electron-ion streaming instabilities can occur because the threshold currents are much smaller and they can occur for $J = 0$. Based on the observed currents the ion-beam-driven instability is the most plausible source of ion-acoustic waves. This result is consistent with the recent observations of \cite{mozer2020}, who concluded that the ion-ion-acoustic instability was the most likely source of ion-acoustic waves. Statistically, we find that only rarely do the waves occur at or within regions of strong current. However, they more frequently occur in turbulent regions of the solar wind where strong current structures are more common. We conclude that the observed waves are likely the result of ongoing turbulence in the solar wind, which can modify the ion or electron distribution functions to generate ion-acoustic waves. Simulations have shown that ion beams can form during solar wind turbulence, which subsequently excite ion-acoustic-like waves \cite[]{valentini2008,valentini2011}. These electrostatic waves may then play a role in energy conversion associated with ongoing turbulence in the solar wind. 

\section{Conclusions} \label{conclusions}
In this paper we have presented observations of ion-acoustic and Langmuir waves observed by Solar Orbiter. We have investigated the association of these waves with currents and current structures in the solar wind around 0.5~AU. The key results are: 

   \begin{enumerate}
      \item Both the LFR and TDS receivers which are part of the RPW instrument onboard Solar Orbiter frequently observe ion-acoustic waves in the solar wind at 0.5~AU. When both the LFR and TDS receivers capture waveform snapshots simultaneously the electric field computed from the probe potentials are nearly identical, which indicates that the onboard processing is reliable for the two receivers.
      \item We compared the frequency of Langmuir waves with the electron plasma frequency calculated from the spacecraft potential. We find that the peak Langmuir wave frequencies typically occur just above the calculated electron plasma frequency. This deviation from the plasma frequency is consistent with the increased frequency above the electron plasma frequency due to thermal corrections to the Langmuir dispersion relation and Doppler shift. In some cases it may be possible to estimate the Langmuir wave number based on this frequency difference. We have provided one example to illustrate this, where multiple Langmuir waves at different frequencies are observed in uniform solar wind conditions. The predicted wave numbers are in agreement with expectations. 
      \item Based on the observed currents in the solar wind, we find that ion-acoustic waves cannot be driven by a simple electron-ion drift instability. Rather complex electron and ion distributions with multiple components are required to generate ion-acoustic waves. The electron-electron-ion streaming instability and the ion-beam driven instability remain possible candidates for generating the ion-acoustic waves, with the ion-beam driven instability being the most plausible.
	\item We find that ion-acoustic waves are observable in the solar wind about $1 \%$ of the time at 0.5 AU. We find that Langmuir and ion-acoustic wave occurrences are associated with solar regions where currents are enhanced. However, the waves typically do not occur at current structures. Rather the waves are typically embedded in extended regions of elevated levels of current occurrences. We propose that the waves are associated with ongoing solar wind turbulence, rather than specific current structures. 
   \end{enumerate}

\begin{acknowledgements}
  	We thank the Solar Orbiter team and instrument PIs for data access and support. Solar Orbiter data are available at http://soar.esac.esa.int/soar/\#home. This work is supported by the Swedish National Space Agency, Grant 128/17. 
\end{acknowledgements}


\end{document}